# Finite-temperature transition-window weights on a fault-tolerant quantum computer: duplex filtering on a thermal purification

Yujin Tang,[1] Yunlong Liu,[1] Miao Xu,[2] and Qing-Xing Xie[1],*

[1]School of Physics, Hubei University, Wuhan 430062, China

[2]Normal College, Hubei University, Wuhan 430062, China

*Corresponding author: xieqx@hubu.edu.cn

## Abstract

The excite–filter–readout framework of Kharazi et al. computes zero-temperature response windows on a fault-tolerant quantum computer from a pure ground state. We extend it to Gibbs states, where the two objects decouple: filters act on absolute energies, a thermal response on transition frequencies. The extension rests on a Liouville-space/TFD identity, explicit in the block-encoding/QSP setting: the thermally weighted transition-window weight is a spectral projection of $H_0 \otimes I - I \otimes H_0^T$ onto the Schmidt (thermofield-double) purification, instance of Rall's framework. Around it we assemble a routed-SELECT block encoding (one Hamiltonian query per filter query), a conjugate-pair readout, and a four-route (duplex, bin, Fourier-LCU, METTS) compatibility analysis. The cost is multiplicative and conditional: a per-window query bound $M_{\mathrm{AE}} \cdot (2 \cdot \sqrt{d/Z_{\mathrm{lb}}} \cdot d_{\mathrm{vt}} + 4 \cdot d_W)$, with a tight signal-adaptive count, the optimal $\sqrt{\alpha\beta}$ degree, and a window-edge term. The preparation factor $\sqrt{d/Z}$ grows exponentially with system size at fixed temperature in positive-entropy-density systems (2.08 → 25.7, N = 8 to 14 at $\beta$ = 2), relieved only as $\beta \to 0$. Zero temperature is recovered for $\beta G \geq \ln(2(d-1)\|D\|^2/\varepsilon)$. Identities are cross-checked between independent formula paths to $10^{-16} - 10^{-12}$ and the readout assembly is validated at matrix level under a block-encoding promise.

## 1. Introduction

### 1.1 Background

Response functions $\chi^{(n)}(\omega_1, \dots, \omega_n)$ encode how a quantum system reacts to external perturbations; their sum-over-states (SOS) form is the bridge between experiment (absorption spectra, Raman profiles, dynamical structure factors) and electronic-structure theory. On a fault-tolerant quantum computer, the excite–filter–readout framework of the base paper [1] computes exactly the SOS content of a frequency window: with the spectrum shifted so that the ground-state energy is $\lambda_0 = 0$,

$\langle\psi_0|D' \cdot \mathbf{1}_{[a,b]}(H_0) \cdot D|\psi_0\rangle = \sum_{\lambda_n \in [a,b]} d'_{0n}\, d_{n0}$ (Eq. 1),

i.e., apply a dipole D, filter with a spectral indicator implemented by quantum signal processing (QSP), apply a second dipole $D'$, and read the non-unitary Hadamard test / amplitude estimation. The construction iterates to arbitrary order, and its query complexity is fully characterized.

The framework, however, is formulated under assumption A4: the initial ensemble is the pure ground state. This is an excellent approximation for the electronic spectra of molecules at room temperature ($\beta_{el} \approx 10^3 Ha^{-1}$), but it fails in the regimes where fault-tolerant quantum simulation is most motivated: doped semiconductors, correlated lattices, Mott systems and warm dense matter, whenever the dimensionless coupling $\beta G$ (inverse temperature × spectral gap) is O(1). The theorems of [1] do not cover the thermal case; we supply that extension here.

## 1.2 The structural problem: absolute energies vs. transition frequencies

A quantum circuit can implement spectral functions of an operator it block-encodes (in the base framework, of $H_0$). The spectral parameter of $\mathbf{1}_{[a,b]}(H_0)$ is the absolute energy $\lambda_n$. The spectral content of a thermal response, however, lives on transition frequencies $\omega_{mn} = \lambda_n - \lambda_m$, because the SOS of the thermal response carries Boltzmann weights $p_m$ on initial states m:

$$\chi_\beta(\omega) = \sum_{m,n} p_m \, |d_{mn}|^2 \, [1/(\omega_{mn} - \omega - i\gamma) + 1/(\omega_{mn} + \omega + i\gamma)].$$

At zero temperature, $p_m \to \delta_{m0}$ and the shift $\lambda_0 = 0$ makes $\lambda_n = \omega_{n0}$: absolute windows and transition windows coincide. This is the role of the $\lambda_0 = 0$ convention in the base paper. At finite temperature the two notions decouple, and the semantics of every window must be stated explicitly (Sec. 3). Reconciling what a circuit can measure with what the physics needs is the technical core of the paper (Sec. 4).

## 1.3 Contributions

The paper delivers a complete, resource-accounted prescription for thermal transition-window weights. Its core is the duplex transition-window identity (Theorem 3): the thermally weighted transition-frequency window weight is the expectation of a spectral projector of $H_\Delta = H_0 \otimes I - I \otimes H_0^T$ on the Schmidt/TFD purification. Three constructions turn the identity into an algorithm: a routed-SELECT block encoding of $H_\Delta$ at one H-family query per filter query; a conjugate-pair readout realizing the real window polynomial in a single QSP sequence; and a four-route compatibility matrix (duplex Δ, bin, Fourier-LCU from any $\rho_\beta$, METTS) recording which preparation feeds which readout. Theorem 10 prices the assembly multiplicatively, with a tight signal-adaptive iterate count, the optimal $\sqrt{\alpha\beta}$ kernel degree, and an explicit ramp term.

The identity has a lineage, which we state once. By the vectorization identity $\mathrm{vec}([H,X]) = (H_0 \otimes I - I \otimes H_0^T)\mathrm{vec}(X)$, $H_\Delta$ is the matrix of the commutator superoperator $\mathcal{L}_H$, and $|\Phi_\beta\rangle \propto \mathrm{vec}(\rho_\beta^{1/2})$, so Theorem 3 is the Hilbert–Schmidt superoperator spectral projection: a standard object of Liouville-space spectroscopy [2] and of the TFD hat-Hamiltonian [3], with classical purification methods using the same doubled structure [4]. It is also implicit in the block-encoding literature: [5] estimates $\mathrm{Tr}[\rho B w(H) C]$-type windows for any $\rho$ with a purification-preparation unitary, and [6] notes the thermal extension. What this paper supplies is the explicit transition-frequency instantiation and its accounting.

Sections 3 and 4 build the thermal identity family and the route landscape: Id-n-β (Lemma 1), sufficiency of $S_\beta^L$, the exact finite-γ dual-weight FDT form $\mathrm{Im}\chi_\beta = S^L(\omega) -$ $S^L(-\omega)$, which holds at finite broadening where the textbook form does not, the expectation-value form as the correct duplex object (cyclic traces compute a density-of-states-type object instead, pinned numerically), the $D' \neq D$ and tensor-component extensions, and the limits of the single-copy alternatives: Cor. 4.1 (no spectrum-independent finite separation rank, while fixed spectra do admit finite representations), the per-initial-state widening of Prop. 4, Prop. 5, and the Liouville observation that higher order reuses the same $H_\Delta$ register. Section 5 treats preparation and readout together: the duplex route requires a Schmidt/TFD preparation (Lemma 7; [7]). Section 7 validates the assembly numerically, through exact diagonalization on Heisenberg rings and Hubbard plaquettes, two independent formula paths, and matrix-level checks of the readout.

## 2. Related work

Thermal-state preparation. The cost of preparing Gibbs states is a classical touchstone: Terhal and DiVincenzo [8] connected thermal preparation to the complexity of partition functions and correlation functions. Poulin and Wocjan [9,10] gave the QPE-plus-amplification family: ground-state preparation in time scaling as the square root of the Hilbert-space dimension [9], and thermal Gibbs sampling with partition-function evaluation whose cost is governed by a universal thermalization-time upper bound $D^\alpha(\alpha < 1/2$ related to the free-energy density) [10]. Quantum Metropolis sampling [11,12] and its variants achieve quadratic mixing speedups [13]; Chen et al. [7] give provably efficient preparation of purified Gibbs states (thermal-field-double type) for rapidly thermalizing systems, with cost $\tilde{O}(\beta \cdot t_{mix}/\varepsilon \cdot t_{mix})$ (incoherent; quoted as in Table 1 of the arXiv v2 of [7], where $t_{mix}$ appears twice) or $\tilde{O}(\beta^2\|H\|/\lambda_{gap}^{3/2} + \beta/(\varepsilon \cdot \lambda_{gap}^{3/2}))$ (coherent; same table). Their coherent algorithm outputs (Thm. I.4, Eqs. (48)–(49)) an approximation to the *canonical purification* $\sum_i e^{-\beta E_i/2} |\psi_i\rangle \otimes |\psi_i^*\rangle/\sqrt{Z}$ in vector norm, which is the Schmidt/TFD structure our duplex readout requires (Sec. 5.1). Rouzé et al. [14] prove that at sufficiently high temperature the dissipative preparation time scales logarithmically in system size; Ding et al. [15] give an end-to-end system–bath analysis; Kuwahara et al. [16] give the 1D thermal area law and a quasi-linear classical algorithm. Our Lemma 7 (direct QSP preparation) is the baseline member of this family with no mixing-time assumption. Which preparation route can feed which readout is a classification, not a free plug-in choice: routes that output a canonical purification within error $\varepsilon$ (route (i) exactly; [7] coherently with $\varepsilon$ in the budget) are duplex-compatible; routes that output a mixed Gibbs state (Metropolis/Lindbladian family [11,13–16]) feed the bin and Fourier-LCU routes (Sec. 5.1).

Thermal typicality. Canonical typicality [17,18] and thermal pure quantum (TPQ) states [19,20] underlie route (ii); quantum-algorithmic versions exist for thermodynamic observables [21,22]. The interplay between typicality's per-state concentration error and amplitude estimation's per-state precision, with total ε-scalings from $\varepsilon^{-2}$ to $\varepsilon^{-3}$ depending on the error allocation (Lemma 8), is, to our knowledge, not previously quantified for spectrally resolved response functions (within our search scope). The estimator-dependent structure of its systematic ensemble bias is quantified here (Lemma 8 and the dual-estimator experiment of Sec. 5.3).

METTS. Minimally entangled typical thermal states [23,24]; the NISQ adaptation AVQMETTS [25]. Our Lemma 9 formalizes METTS response readout as an *adaptive-bin* version of the zero-temperature circuit and states explicitly what the entropy $e^{S(\pi_\beta)}$ does and does not control (the sampling theory is not supplied; Sec. 5.4).

Finite-temperature response on quantum hardware. Kökcü et al. [26] (ancilla-free linear-response framework, NISQ demonstration), Sun et al. [27] (QITE + real-time, finite-T dynamical correlations), Abhishek et al. [28] (nuclear linear response), Bishop et al. [29] (Fermi–Hubbard Green functions, zero-T setting), Selisko et al. [30] (DMFT impurity solver via Lehmann representation), Dhawan et al. [31] (imaginary-time Green functions, analytic continuation), Irmejs and Santos [32] (constant-depth Krylov correlation functions). On the fault-tolerant side, Loaiza et al. [33] estimate nonlinear spectroscopy via generalized QPE: their formalism admits any equilibrium input state commuting with H (ground, excited, or thermal, their Sec. II.1), but the GQPE primitive consumes energy eigenstates and thermal weights enter only as classical coefficients $\rho_n$; no transition-window filtering identity is given, and no β-parameterized query complexity. The zero-temperature excite–filter–readout framework itself is the base paper [1].

Response estimation in the block-encoding framework; the Liouville/TFD form of Theorem 3. Rall [5] gives block-encoding algorithms for n-time correlation functions, densities of states, and dynamical linear response functions, with amplitude estimation reaching the quadratic $(1/\varepsilon)$ accuracy scaling; his framework is modular in the state preparation, applies to any $\rho$ given a purification-preparation unitary (thermal states included), and his Lemma 4 prepares thermal states by applying $e^{-\beta H/2}$ to a purification of the maximally mixed state with amplitude amplification at $\sqrt{D/Z}$ cost, the same construction and scaling as our Lemma 7. Roggero and Carlson [6] give a zero-temperature dynamic linear-response algorithm with controlled errors and note that preparing thermal equilibrium states extends the framework to finite temperature. The vectorized/Liouville-space form of our Theorem 3, $S_\beta^W = \langle\langle D\rho^{1/2}|\mathbf{1}_W(\mathcal{L}_H)|D\rho^{1/2}\rangle\rangle$ with $\mathcal{L}_H = [H,\cdot]$, via vec([H,X]) = $(H \otimes I - I\otimes H^T)$vec(X) and $|\Phi_\beta\rangle \propto vec(\rho^{1/2})$, is standard: the TFD hat-Hamiltonian $H - \tilde{H}$ is the time-translation generator of the doubled space [3], Liouville-space superoperator spectroscopy is the classical formalism for exactly this object [2], and classical purification-based finite-temperature dynamics (e.g. [4]) use the same doubled structure. What remains specific to us relative to [5,6] is the transition-frequency instantiation: the routed-SELECT block encoding of the difference operator, the conjugate-pair real-polynomial readout with its leakage analysis, the preparation–readout compatibility analysis, and the multiplicative resource accounting of Theorem 10. Conversely, [5] covers response quantities of a single Hamiltonian at a generality (arbitrary B, C, window functions w; density-of-states sketches) that this paper does not attempt to match; the instantiation here is complementary, not broader.

TFD states on quantum computers (preparation line). Miceli and McGuigan [34] implement thermo-field dynamics on a quantum computer via temperature-encoded Hamiltonians (their VQE attempt fails as a basis change; the thermo-double state is prepared and measured; no difference operator, no response readout); Zhu et al. [35] prepare TFD states of the transverse-field Ising model variationally on an ion trap (benchmarked by equal-time one- and two-body correlators, both intra-system and inter-system); Wu et al. [36] give the variational TFD thermal-simulation protocol. [34] and [35] are checked at full-text level, [36] at abstract level (its preparation-only content is corroborated by [35]'s full text); none of the three measures response functions or filters transition frequencies. On the preparation-theory side, the modular-annihilator parent-Hamiltonian construction for purified Gibbs states [37] resolves transition frequencies for parent-Hamiltonian spectral design, preparation-adjacent, with no readout algorithm; it is listed in `survey_queries.md`.

Table 1. Method-by-method comparison (five elements). The nearest methods, classified by input state, filtered operator, measured estimator, window semantics, and query model:

| method | input state | operator filtered | measured estimator | window semantics | query model |
|---|---|---|---|---|---|
| base framework [1] | pure ground state $\|\psi_0\rangle$ (A1: overlap $p_0$, gap G) | absolute-energy indicator $\mathbf{1}_{[a,b]}(H_0)$ | $\langle\psi_0\|D'\mathbf{1}_{[a,b]}(H_0)D\|$ı (QSP + Hadamard/AE) | absolute window on $\lambda$ | $U_H$ queries, $\tilde{O}((\alpha/\gamma_w)\cdot\zeta_\omega/(\varepsilon_{\mathrm{rel}}\cdot\mu))$ per window, prep additive (persistent-state convention) |
| GQPE spectroscopy [33] | energy eigenstate $\|\varepsilon_{n_0}\rangle$ per run; thermal $\rho_0$ via classical sampling of $U_{\varepsilon_{n0}}$ (or Stinespring dilation of $\rho_0$) | none — time registers in a window-function superposition; frequencies read out by QFT | $R(\tilde{\omega})$ (multi-variate correlation, Lehmann-sum content); AE at a point or Bernoulli sampling of $\|R\|^2$ | frequency content via time-domain resolution $\Delta$ | $\tilde{O}(D\cdot\Omega/(\eta\cdot\varepsilon))$ time-evolution queries per point (their Thm. 1: $\Omega$ = product of operator 1-norms, $\eta$ = energy resolution, $\varepsilon$ = accuracy; an $\varepsilon^{-4}$ variant samples all points simultaneously ) |
| coherent thermal prep [7] | — (preparation only) | — | canonical purification $\|\rho_\beta^{1/2}\rangle$ to vector-norm error $\varepsilon$ | — | $\tilde{O}(\beta^2\|H\|/\lambda_{gap}^{3/2}+\beta/(\varepsilon\cdot\lambda_{gap}^{3/2}))$ per prep call |
| block-encoding quantities [5] | any $\rho$ with a purification-preparation unitary (thermal included; Lemma 4 ≈ our Lemma 7) | w(H) of a single given Hamiltonian (KPM sketches, LDOS, linear response) | $\mathrm{Tr}[\rho B w(H) C]$ via amplitude estimation | absolute window on $\lambda(H)$ | $\tilde{O}(1/\varepsilon)$ AE scaling; preparation modular (black-box) |
| linear-response framework [26] | arbitrary (NISQ; ancilla-free) | none (time-domain correlators) | $\mathrm{Tr}[\rho D(t) D]$ type correlators | none (time series) | not a fault-tolerant query theorem |
| this paper (Thm. 3 + 10) | Schmidt/TFD purification $\|\Phi_\beta\rangle$ (route-dependent prep) | difference-operator indicator $\mathbf{1}_W(H_\Delta), H_\Delta = H_0\otimes I - I\otimes H_0^T$ | $\langle\Phi_\beta\|(D\otimes I)\mathbf{1}_W(H_\Delta)(D\otimes I)\|\Phi_\beta\rangle\equiv S_\beta^W$ as an identity; estimated to $\varepsilon_{\mathrm{rel}}$ (exact-reflection AE) | transition window on $\omega_{mn}=\lambda_n-\lambda_m$ | $U_H$-family queries; single Õ expression with multiplicative preparation (Theorem 10) |

Positioning (search-scope records in `survey_queries.md`). Theorem 3 is the transition-frequency instance of the [5] framework: system = two copies, Hamiltonian = $H_0\otimes I-$ $I \otimes H_0^T$, state = canonical purification; an instance that [5,6] do not work out, and whose value is that it makes the *transition-frequency* semantics (the axis the thermal response actually lives on) the circuit's native axis. Relative to [33], the closest fault-tolerant spectroscopy work: their primitive consumes eigenstates with classical thermal weights; ours filters transition frequencies coherently on a purification. Relative to the TFD-preparation line [34–36]: those works establish that TFD states are preparable on quantum hardware but measure no response objects; our Lemma 7 and the compatibility analysis of Sec. 5.1 connect that preparation class to the duplex readout. The architecture-level claim, routed-SELECT difference-operator block encoding at one H-family query per filter query, conjugate-pair real-polynomial readout, four-route compatibility matrix, and the multiplicative accounting of Theorem 10, is, within our documented search scope, new.

---

# 3. Setup and thermal identities

## 3.1 Conventions

$H_0$ has eigenpairs $(\lambda_j, |\psi_j\rangle)$ with the spectrum shifted so that $\lambda_0 = 0$ (the base-paper convention; all windows below live in the shifted $\lambda$ coordinate, "measured from the ground-state energy"). The thermal state is $\rho_\beta = e^{-\beta H_0}/Z$ with Z = $\sum_j e^{-\beta\lambda_j}$; Boltzmann weights $p_m = e^{-\beta\lambda_m}/Z$ are invariant under energy shifts in ratio form (the numerical value of Z changes by $e^{\beta E_0}$; wherever Z enters numerically (e.g. $\sqrt{d/Z}$) we work in the shifted coordinate). We write $\omega_{mn} \equiv \lambda_n - \lambda_m$. $D, D'$ are Hermitian system operators (dipole, spin, density) with matrix elements $d_{mn} = \langle m|D|n\rangle$; $\alpha_D$ denotes the dipole block-encoding subnormalization (chosen so that $\beta$ always means the inverse temperature). Atomic units; $k_B = 1$. Conjugation conventions (fixed in the computational basis). $H^T$ is the matrix transpose of H in the computational basis and $\bar{H}$ the entrywise complex conjugate; for Hermitian H these coincide $(H^T = \bar{H})$. $|\psi_j^*\rangle$ is the componentwise conjugate of the eigenvector's coordinate vector in that basis, hence $H^T|\psi_j^*\rangle = \lambda_j|\psi_j^*\rangle$: the conjugated copy is an eigenvector of the transposed Hamiltonian. Hermiticity of D and $D'$ (used by Theorem 3 and the $D' \neq$ D extension) is likewise a statement in the computational basis. Purification convention: the energy-basis Schmidt/TFD purification is $|\Phi_\beta\rangle = \sum_j \sqrt{p_j}\, |\psi_j\rangle \otimes |\psi_j^*\rangle$, where * denotes componentwise complex conjugation in the computational basis (the energy-basis expansion then holds with the conjugated eigenvectors, consistently with the transposed-Hamiltonian statement at the end of this paragraph); for Hamiltonians with real eigenvectors (all numerical systems of this paper) the star is vacuous, and the convention is basis-independent in the sense established for canonical purifications (see [7], footnote 25). Equivalently, the second (conjugated) copy carries the transposed Hamiltonian $H^T$ in the computational basis (a full notation index is provided in App. A.0). Two widths are kept distinct throughout: $\gamma$ is a *physical* phenomenological broadening (Prop. 2), $\gamma_w$ *a filter* window width (Theorem 10); they are chosen independently, and shrinking $\gamma$ does not by itself sharpen any QSP window.

Window semantics (endpoint convention). All indicator windows are closed intervals [a,b] evaluated with a 1e-9 tolerance; a spectral line lying within tolerance of an edge is counted *inside* (the numerical `in_window` of the archived scripts), and the sensitivity of any quoted weight to this convention is noted where it arises (the T5h edge-coincident line). $\mathbf{1}_{[a,b]}(\lambda_n)$ is an *absolute* window; $\mathbf{1}_{[a,b]}(\omega_{mn})$ is a *transition* window. The central object of this paper is the thermal transition-window weight

$S_\beta^{[a,b]} \equiv \sum_{m,n} p_m\ |\langle m|D|n\rangle|^2 \cdot \mathbf{1}_{[a,b]}(\lambda_n - \lambda_m)$ (Eq. 2),

the finite-temperature analogue of the base-paper Eq. (5). A second, distinct object enters the typicality route (Sec. 5.3): the absolute-window weight $J_\beta^{[a,b]} \equiv$ $\mathrm{Tr}[\rho_\beta D\mathbf{1}_{[a,b]}(H_0)D] = \sum_{m,n} p_m\ |d_{mn}|^2 \cdot \mathbf{1}_{[a,b]}(\lambda_n)$, a bin-route quantity with different window semantics; on Hei8 at $\beta = 2$, $J_\beta^{[0,3]} = 5.7025$ while $S_\beta^{[1,38]} = 1.0809$: the two must never be conflated, substituting one for the other changes the typicality repetition factor M by $(J/S)^2 \approx 28$ at this anchor (Sec. 5.3 and 5.5). Fig. 1 displays both axes for the 8-site Heisenberg ring: the same interval [1,38] selects different physical content on the two axes, and only the transition axis is the response-theoretic one.

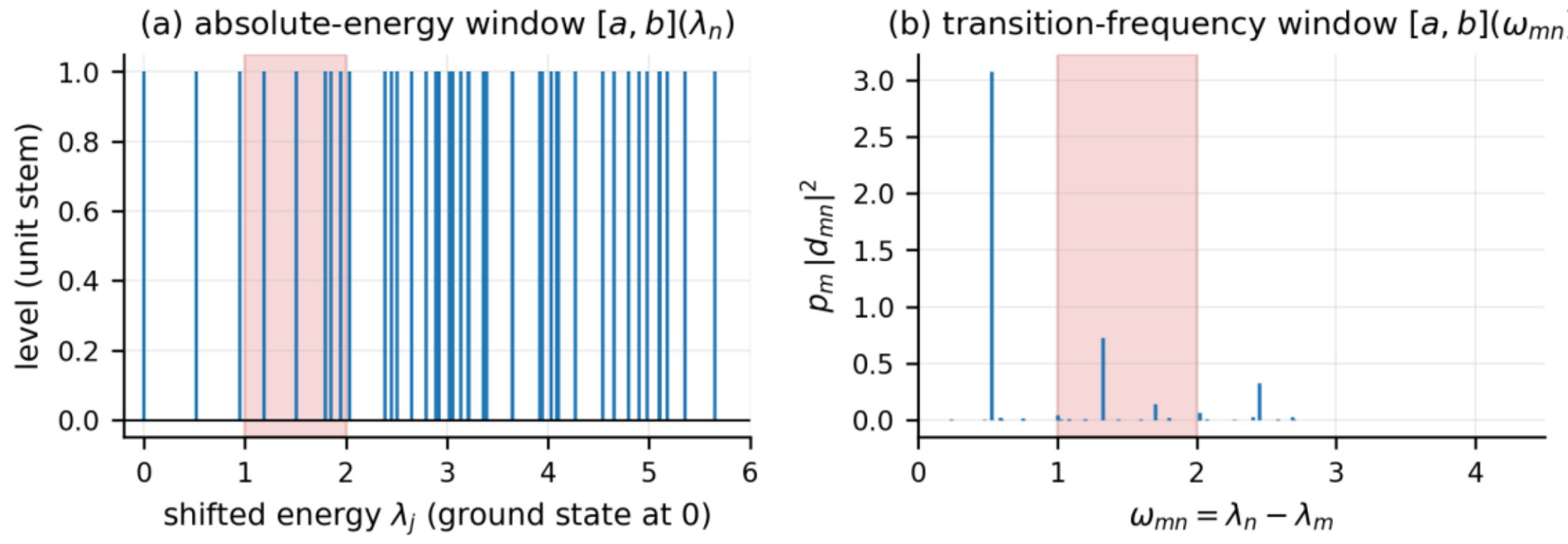

*Fig. 1. Absolute-energy windows vs. transition-frequency windows. (a) The shifted spectrum $\lambda_j$ of the 8-site Heisenberg ring (Sz = 0 sector,* dim *70) drawn as unit stems on the true energy axis; the shaded band is the absolute window* $[1,2](\lambda)$. *(b) The positive-frequency transition lines $\omega_{mn} = \lambda_n - \lambda_m$ weighted by $p_m|d_{mn}|^2$ at $\beta = 2$ (top 80 lines by weight; D = $S^z_{q=\pi}$); the shaded band is the same interval as a transition window* $[1,2](\omega)$. *At zero temperature panel (b) collapses onto panel (a) with m = 0; at finite temperature the two window semantics decouple.*

## 3.2 The thermal nested identity

Lemma 1 ($Id - n - \beta$). For any operator chain $A_0, \ldots, A_n$ and spectral functions $f_1, \ldots, f_n$ (defined on the shifted spectrum),

$\mathrm{Tr}[\rho_\beta A_0 f_1(H_0) A_1 f_2(H_0) A_2 \cdots f_n(H_0) A_n] = \sum_{j_0,\ldots,j_n}$
$p_{j_0}\ f_1(\lambda_{j_1}) \cdots f_n(\lambda_{j_n}) \langle j_0|A_0|j_1\rangle\langle j_1|A_1|j_2\rangle \cdots \langle j_n|A_n|j_0\rangle$. $(Id - n - \beta)$

*Proof.* Expand the trace in the energy eigenbasis and insert n+1 resolutions of the identity; $f_k(H_0)|\psi_j\rangle = f_k(\lambda_j)|\psi_j\rangle$ factors out; the Boltzmann factor is the diagonal matrix element $\langle j_0|\rho_\beta|j_0\rangle = p_{j_0}$. (Full proof: App. A.1.)

$\beta \to \infty$ sends $p_{j_0} \to \delta_{j_0,0}$ and cuts the cycle open at $\langle\psi_0|$, recovering the zero-temperature nested identity of the base paper (Eq. (27) there). The spectral slots $f_k$ hang on absolute energies; at zero temperature the physical denominators $(\lambda_{j_k} - \omega_k - i\gamma)$ are absolute, so the circuit object *is* the physical object; at finite temperature the physical SOS denominators are transition frequencies ($\lambda_{j_k} - \lambda_{j_{k-1}} - \omega_k - i\gamma$), and the mismatch is the subject of Sec. 4.

### 3.3 The thermal linear response and the dual-weight FDT form

Definitions and sign conventions (time domain → Lehmann; one closed chain). All response formulas of this paper follow one convention chain, stated once here. Start from the phenomenologically damped retarded Kubo response ($\hbar = 1$):

$\chi_{Kubo}(t) = -i \cdot \theta(t) \cdot e^{-\gamma t} \cdot \langle [D(t), D(0)] \rangle_\beta$, D(t) = $e^{iH_0 t} D e^{-iH_0 t}, \gamma > 0$,

with Fourier transform convention ∫₀^∞ dt $e^{i\omega t}$. Expanding $C^>(t) = \langle D(t)D(0)\rangle = \sum_{m,n} p_m\, |d_{mn}|^2 e^{-i\omega_{mn} t}$ and $C^<(t) = \langle D(0)D(t)\rangle = \sum_{m,n} p_m\, |d_{mn}|^2 e^{+i\omega_{mn} t}$ and integrating term by term ($\int_0^\infty e^{i(\omega \mp \omega_{mn})t}\, e^{-\gamma t} dt = i/(\omega \mp \omega_{mn} + i\gamma)$) gives

$$\chi_{Kubo}(\omega) = -\sum_{m,n} p_m\, |d_{mn}|^2 \cdot [1/(\omega_{mn} - \omega - i\gamma) + 1/(\omega_{mn} + \omega + i\gamma)].$$

This paper's $\chi_\beta$ is defined as the negative of that object, $\chi_\beta \equiv -\chi_{Kubo}$; the absorptive-polarizability convention, under which $\mathrm{Im}\chi_\beta = S^L(\omega) - S^L(-\omega)$ is positive on absorption peaks (equivalently, $\chi_\beta(t) = +i \cdot \theta(t) \cdot e^{-\gamma t} \cdot \langle [D(t), D(0)] \rangle_\beta$ with the same transform; both forms of the definition are used interchangeably below). Every response statement in this paper refers to this $\chi_\beta$; readers who prefer the Kubo sign should flip the overall sign of $\chi_\beta$ throughout, leaving Im-related identities (FDT-L) unchanged in magnitude. With this definition,

$$\chi_\beta(\omega) = \sum_{m,n} p_m\, |d_{mn}|^2 \cdot [1/(\omega_{mn} - \omega - i\gamma) + 1/(\omega_{mn} + \omega + i\gamma)].$$

The two Lehmann terms add; the sign convention is the one fixed at the head of this section (the numerical three-way check and the two-level closed form are in App. H).

Proposition 2. Let $\chi_\beta(\omega)$ be the damped Kubo response of D under the convention chain above (standard theory [38]). Then: (a) *(Lehmann, two-term form)* $\chi_\beta(\omega) = \sum_{m,n} p_m\, |d_{mn}|^2 [1/(\omega_{mn} - \omega - i\gamma) + 1/(\omega_{mn} + \omega + i\gamma)]$, the two Lehmann terms add; the sign matters, because $\mathrm{Im}[1/(\omega_{mn} + \omega + i\gamma)] = -L_\gamma(\omega_{mn} + \omega)$ already carries the minus that (FDT-L) needs (a minus *between* the terms would give $\mathrm{Im}\chi = S^L(\omega) + S^L(-\omega)$; App. A.2, sign verified numerically in the supplemental sign check). (b) *(dual-weight FDT)* With the thermal Lorentzian spectrum $S^L_\beta(\omega) \equiv \sum_{m,n} p_m\, |d_{mn}|^2 L_\gamma(\omega_{mn} - \omega)$, $L_\gamma(x) = \gamma/(x^2 + \gamma^2)$:

$\mathrm{Im}\chi_\beta(\omega) = S^L_\beta(\omega) - S^L_\beta(-\omega) = \sum_{m,n} (p_m - p_n) |d_{mn}|^2 L_\gamma(\omega_{mn} - \omega)$ (FDT-L),

exactly, at finite $\gamma$; the second equality follows by the dummy-index exchange ($m \leftrightarrow n$); a pure relabeling, with no $e^{-\beta\omega_{mn}}$ weight substitution (that substitution does not hold for the broadened spectrum; the numerical check V4 of Table E.1 demonstrates it). (c) *(textbook form; sufficient conditions, and where it fails)* The distribution-level identity $\mathrm{Im}\chi_\beta = (1 - e^{-\beta\omega}) \cdot S_\beta$ holds on δ-support, with $S_\beta = \pi^{-1}\cdot$(the unbroadened positive- and negative-frequency spectral measure); equivalently $S^L_\beta$ in the limit $(1/\pi)L_\gamma \to \delta$. At finite $\gamma$ the pointwise form generally fails when the textbook factor multiplies the *broadened* $S^L_\beta$: the Lorentzian convolution and the multiplication by the frequency-dependent factor $(1 - e^{-\beta\omega})$ do not commute. Sufficient conditions for the pointwise form at finite $\gamma$ (App. A.2(c)): (i) the δ-support limit; (ii) *isolated lines*, at the peak position $\omega = \omega_{mn}$ of a transition separated by $\delta_{line}$ from every other thermally weighted line, with $\gamma \ll \delta_{line}$ (tail contamination ≤ $2\mathrm{Tr}[\rho_\beta D^2] \cdot \gamma/\delta^2_{line}$) and $\beta\gamma \ll 1$ (the frequency factor varies by $O(\beta\gamma)$ across the peak); (iii) special spectra with accidental (e.g. symmetry-forced) cancellations are not excluded; the conditions are sufficient, not necessary, and the dense-Lorentzian measurements of Fig. 3 demonstrate non-necessity only. At finite $\gamma$ the tails are present: measured deviations up to 1.0002 (absolute) / 0.292 (relative) at $\gamma = 0.05$ and up to 0.477 (absolute) at $\gamma = 0.01$ on the 8-site ring (Fig. 3; deviation = $|\mathrm{Im}\ \chi_{\text{two-term}} - (1 - e^{-\beta\omega})S^L(\omega)|$, maximum over the 21-point grid $\omega \in [0.2, 3.0]$ of the V4 scan) are this tail contamination, not a failure of (FDT-L). (d) *(sufficiency)* $S^L_\beta$ on the full $\omega$ axis determines $\chi_\beta(\omega)$ completely: both Lehmann terms are explicit functionals of $S^L$ (the second window is the mirror of the first).

*Proof sketch.* (a) standard two-term Kubo–Lehmann form; (b) take imaginary parts and relabel $(m \leftrightarrow n)$; (c) at finite $\gamma$ the Lorentzian tails break the detailed-balance substitution except on δ-support or at isolated lines; (d) both Lehmann terms are ω-rescalings of $S^L$. Full derivation: App. H (App. A.2).

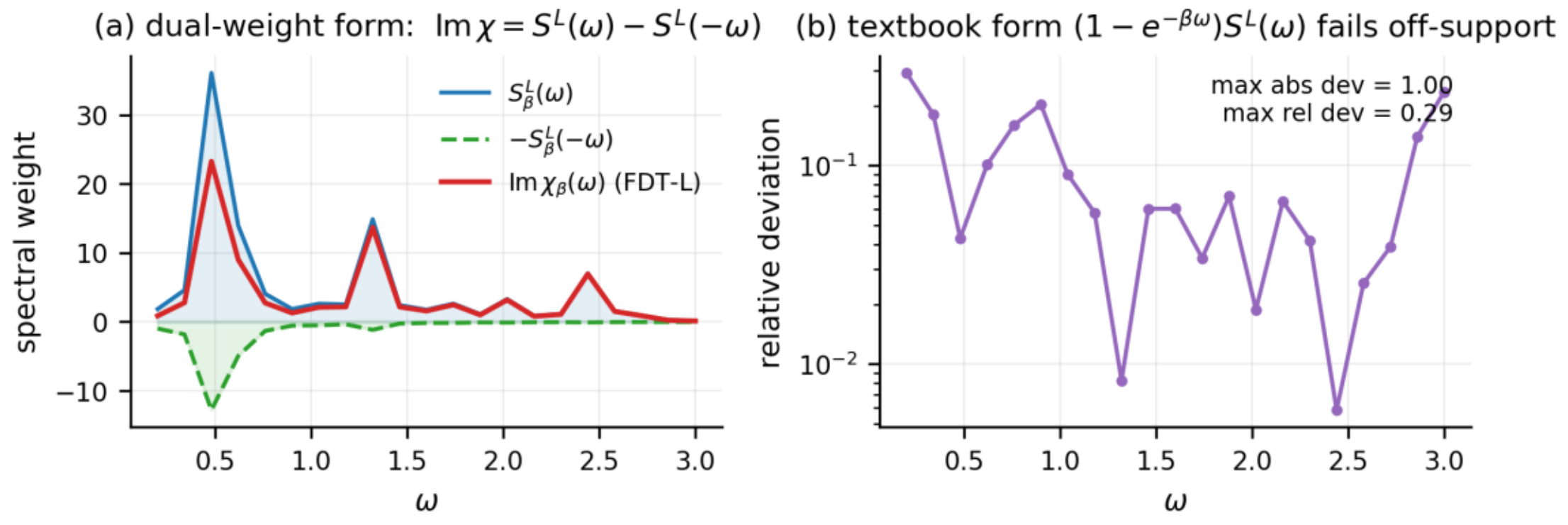

*Fig. 2. The dual-weight fluctuation–dissipation form (FDT-L); cross-checked against the independent evaluation of V4 to* $7.5 \times 10^{-14}$*. Heisenberg ring N = 8,* $\beta = 2, \gamma = 0.05, D = S^z_{q=\pi}$*. (a) The thermal Lorentzian spectrum* $S^L_\beta(\omega)$ *(blue, filled), its negative-frequency mirror* $-S^L_\beta(-\omega)$ *(green dashed, filled), and the response* $\mathrm{Im}\chi_\beta = S^L(\omega) - S^L(-\omega)$ *(red). The mirror window is measured by the same circuit with the mirrored frequency window, at zero extra design; if the mirrored window is in the measurement plan, the consistency check costs zero additional queries (Cor. 10.2). (b) Pointwise relative deviation of the textbook form* $(1 - e^{-\beta\omega})S^L(\omega)$ *from the exact two-term* $\mathrm{Im}\chi$ *(Prop. 2(a), plus-sign Lehmann form): up to 0.29 relative (1.00 absolute), the quantitative signature of Lorentzian-tail contamination at finite* $\gamma$*.*

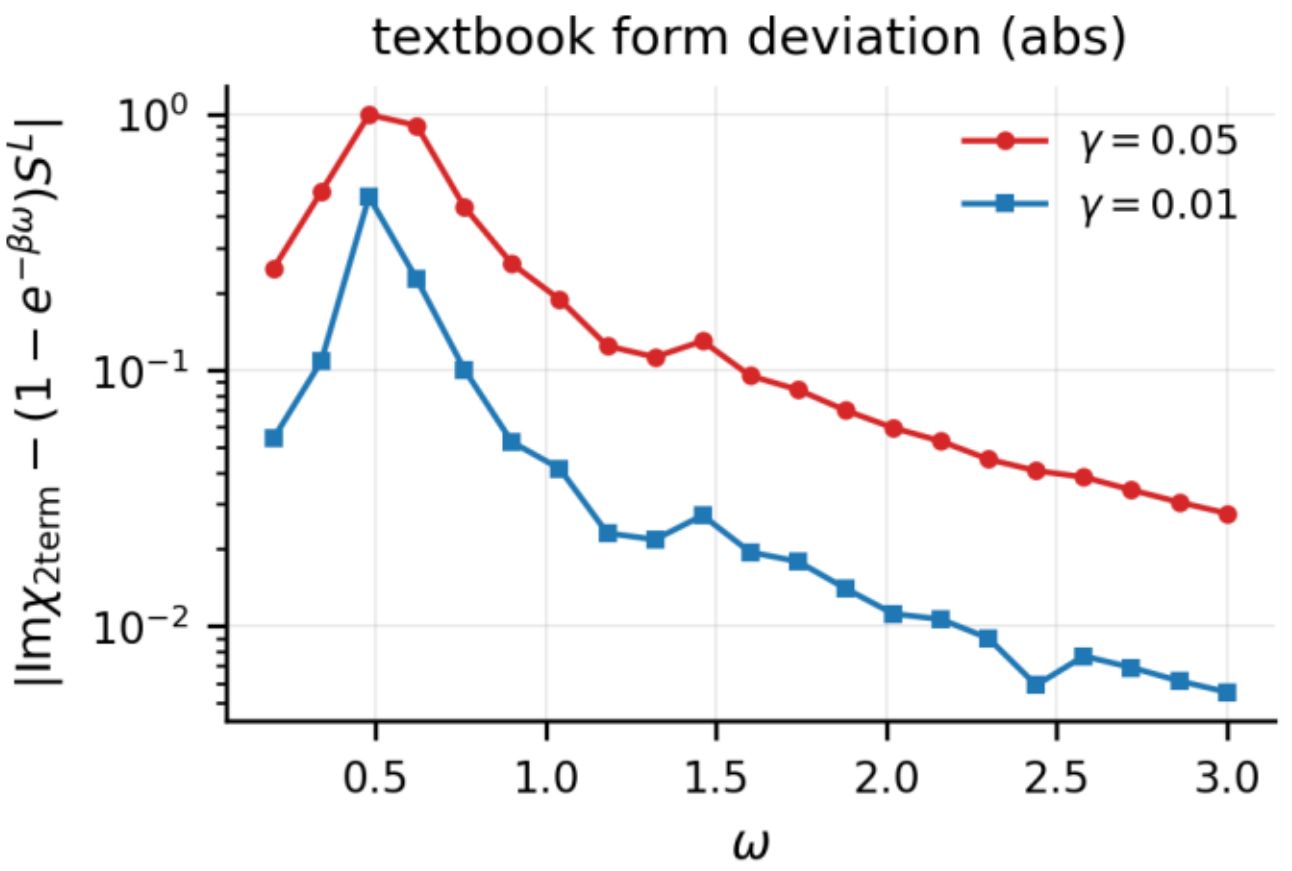

*Fig. 3. Textbook-form deviation vs. broadening. Absolute deviation $|\mathrm{Im}\,\chi_{two-term} - (1 - e^{-\beta\omega})S^L(\omega)|$ on the $\omega$ grid for $\gamma = 0.05$ and $\gamma = 0.01$ (Hei8, β = 2). The deviation falls monotonically as $\gamma$ shrinks (1.0002 at $\gamma = 0.05$ vs. 0.4775 at $\gamma = 0.01$ at the strongest grid point) but not linearly in $\gamma$ (peak-region deviations behave like a tail-rearrangement term) consistent with δ-support / isolated-line-only validity.*

Algorithmic meaning. The only object the quantum side must estimate is $S^L_\beta(\omega)$, or, for window estimation, its window aggregate. Two window objects must be kept apart: the Lorentz-window aggregate $S^{L,W}_\beta = \sum p_m |d_{mn}|^2 \cdot (L_\gamma * \mathbf{1}_W)(\omega_{mn})$ (the QSP-realizable smoothing of a width-$\gamma_w$ indicator, $\gamma_w$ *a filter* width chosen independently of any physical broadening $\gamma$), and the sharp indicator-window weight $S^{[a,b]}_\beta$ of Eq. 2, which the QSP realization approximates to $\varepsilon'$ off the transition band. The algorithmic deliverable is the estimation of these window weights (Theorem 10). Completeness (Prop. 2(d): the full Lorentzian spectrum determines $\chi_\beta$) is a statement about information content, not an algorithm, and the pointwise response does not require an inverse problem on window data: the Lorentz kernel $L_\gamma(H_\Delta - \omega)$ and the real-part kernel $(H_\Delta - \omega)/((H_\Delta - \omega)^2 + \gamma^2)$ are themselves QSP-realizable functions of the same doubled register, so the pointwise damped $\mathrm{Im}\chi_\beta$ and $\mathrm{Re}\chi_\beta$ are natural follow-up primitives of this machinery (Sec. 8.2 (iv)); what we do not supply is the adaptive protocol (window widths/numbers against unknown line spacing) and the per-point cost accounting for those kernels. What the window-weight data set provides directly: the imaginary part by mirror subtraction when the mirrored windows are themselves measured (their QSP and discretization errors then enter; Cor. 10.2), and two self-consistency checks on the same data, weight-level detailed balance $S(-\omega) = e^{-\beta\omega}S(\omega)$ (δ-support) and (FDT-L) (Sec. 6, Cor. 10.2). The zero-temperature limit of (FDT-L) reproduces the base paper's $\mathrm{Im}\alpha^{(1)} = L(\omega) - L(-\omega)$ term-by-term (the base paper's Eq. (54)), with the negative-frequency mirror playing the role of the reversed term.

---

## 4. Filtering on transition frequencies: the route landscape

### 4.1 The duplex difference operator (route Δ)

Theorem 3 (duplex transition-energy filtering). Let $|\Phi_\beta\rangle = \sum_j \sqrt{p_j}\,|\psi_j\rangle \otimes |\psi_j^*\rangle$ (energy-basis Schmidt/TFD purification of $\rho_\beta$ on two copies; conjugation convention of Sec. 3.1),

$H_\Delta \equiv H_0 \otimes I - I \otimes H_0^T$ (the second copy carries the transposed Hamiltonian; $H_0$ Hermitian gives $H_0^T = \bar{H}_0$, and $H_0^T = H_0$ for real $H_0$). Then for any window W = [a,b] and $D' = D$:

$$S_\beta^{[a,b]} = \langle \Phi_\beta | (D \otimes I) \cdot \mathbf{1}_{[a,b]}(H_\Delta) \cdot (D \otimes I) | \Phi_\beta \rangle. \text{ (T3)}$$

*Proof.* $(D \otimes I)|\Phi_\beta\rangle = \sum_{j,n} \sqrt{p_j}\, d_{nj} |\psi_n\rangle \otimes |\psi_j^*\rangle$ with $d_{nj} = \langle n|D|j\rangle$; $H_0|\psi_j\rangle = \lambda_j|\psi_j\rangle$ implies by componentwise conjugation $\bar{H}_0|\psi_j^*\rangle = \lambda_j|\psi_j^*\rangle$, i.e. $H_0^T|\psi_j^*\rangle = \lambda_j|\psi_j^*\rangle$, so $H_\Delta|\psi_n\rangle \otimes |\psi_j^*\rangle = (\lambda_n - \lambda_j)|\psi_n\rangle \otimes |\psi_j^*\rangle$ and the projector contributes $\mathbf{1}_{[a,b]}(\lambda_n - \lambda_j)$. The bra side must be expanded with the dipole absorbed; $\langle\Phi_\beta|$ itself is the *single* sum $\sum_r \sqrt{p_r}\, \langle\psi_r| \otimes \langle\psi_r^*|$; it is $\langle\Phi_\beta|(D \otimes I) = \sum_{r,s} \sqrt{p_s}\, d_{rs}^* \langle\psi_r| \otimes \langle\psi_s^*|$ that carries two indices (D Hermitian: $d_{rs}^* = d_{sr}$). Contracting the two dipole-carrying sides forces $\delta_{rn}\delta_{sj}$ and leaves $\sum_{n,j} p_j\, |d_{nj}|^2 \mathbf{1}_{[a,b]}(\lambda_n - \lambda_j) = S_\beta^{[a,b]}$. (Full proof: App. A.3.)

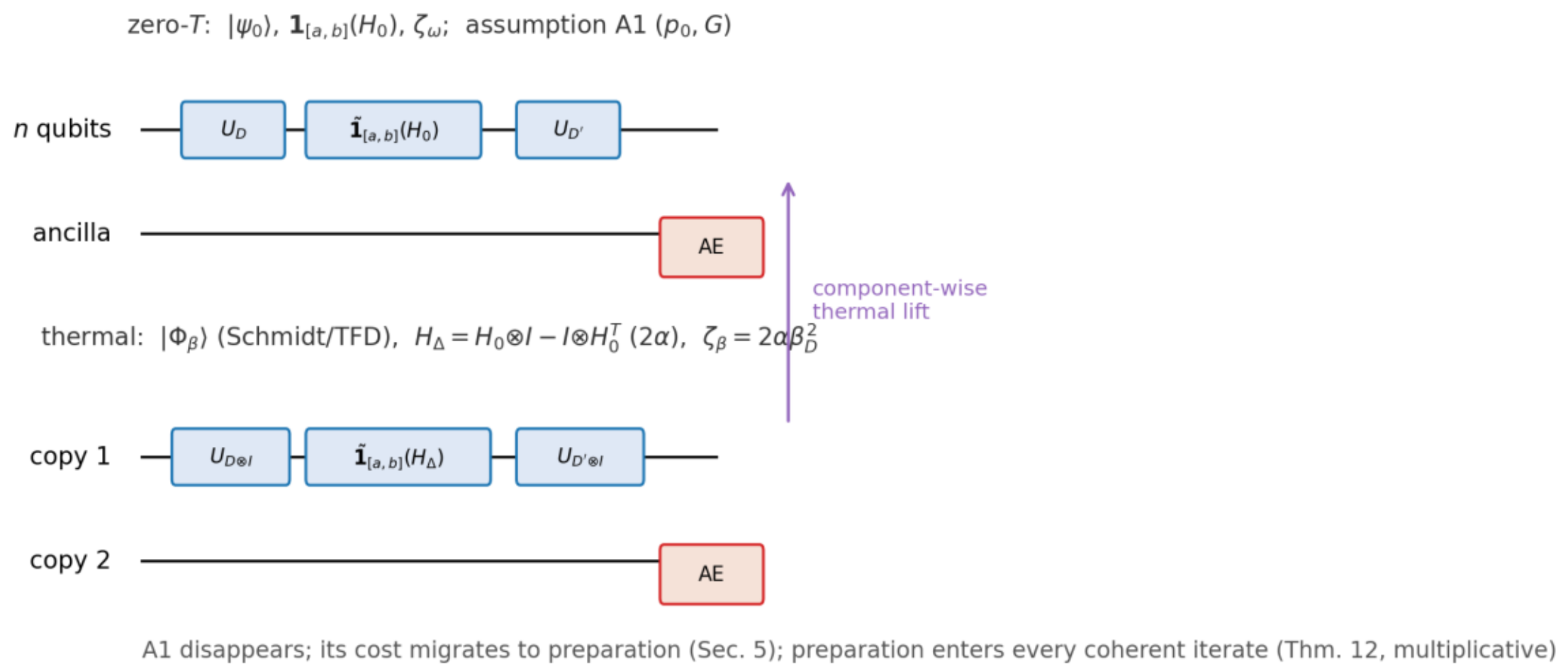

*Fig. 4. Component-wise thermal lift of the excite–filter–readout circuit (schematic). Top: zero-temperature circuit of the base paper [1]. Bottom: thermal duplex circuit of Theorem 3: the initial state becomes the Schmidt/TFD purification $|\Phi_\beta\rangle$; the filter becomes the indicator of the difference operator $H_\Delta = H_0 \otimes I - I \otimes H_0^T$ with block-encoding subnormalization $2\alpha$ (a routed two-term LCU, one Hamiltonian block-encoding query per $H_\Delta$ query; Sec. 4.1); the composite readout subnormalization becomes $\zeta_\beta = 2\alpha\alpha_D^2$; the readout machine (exact-reflection amplitude estimation, App. C) applies the conjugate-pair window realization (two window sequences per chain) per coherent iterate, with the preparation unitary likewise entering every iterate (Theorem 10, multiplicative accounting). The zero-temperature assumptions A1 (overlap $p_0$, gap G) are absent from the readout; their cost migrates to the preparation side (Sec. 5).*

Remark 3.1 (the trace form evaluates a different object). The natural correct object is the expectation-value form $\langle\Phi_\beta|X|\Phi_\beta\rangle, X = (D \otimes I)\mathbf{1}_W(H_\Delta)(D \otimes I)$. Under the cyclic trace, $\mathrm{Tr}[(\rho^{1/2} \otimes I)X(\rho^{1/2} \otimes I)] \equiv \mathrm{Tr}[(\rho_\beta \otimes I)X] = \sum_{m,k,n} p_m\, |d_{km}|^2 \mathbf{1}_W(\lambda_k - \lambda_n)$, a *density-of-states-type* object with a free sum on the second copy, not the spectral weight. On hub22 ($\beta = 2, W = [0.5,1.5]$): S = $5.097 \times 10^{-4}$ while both trace forms equal 3.073865 (identical by cyclicity; equal to the directly computed density-of-states object to 15 digits; full precision in the archived data); see Fig. 5. This is a common source of confusion in two-copy thermal constructions; we pin it with a permanent numerical probe (V3b). A natural single-copy trace form does exist, however, and should not be conflated with the trap above: the time-domain/Fourier-LCU disassembly $S_\beta^W \approx \sum_k c_k \mathrm{Tr}[\rho_\beta e^{iHt_k} D\, e^{-iHt_k} D] = \mathrm{Tr}[\rho_\beta \cdot g(ad_H)(D) \cdot D]$ (g a smoothing of the window indicator; $c_k$ its Fourier coefficients) is preparable on any $\rho_\beta$ realization at a query cost of the same order as the duplex route, the Fourier-LCU route of Sec. 5.1. The duplex route's advantage over it is the sharp-window one-step readout (a width-$\gamma_w$ indicator window realized in a single QSP sequence), not feasibility; an absolute "no single-trace form exists" statement would be wrong.

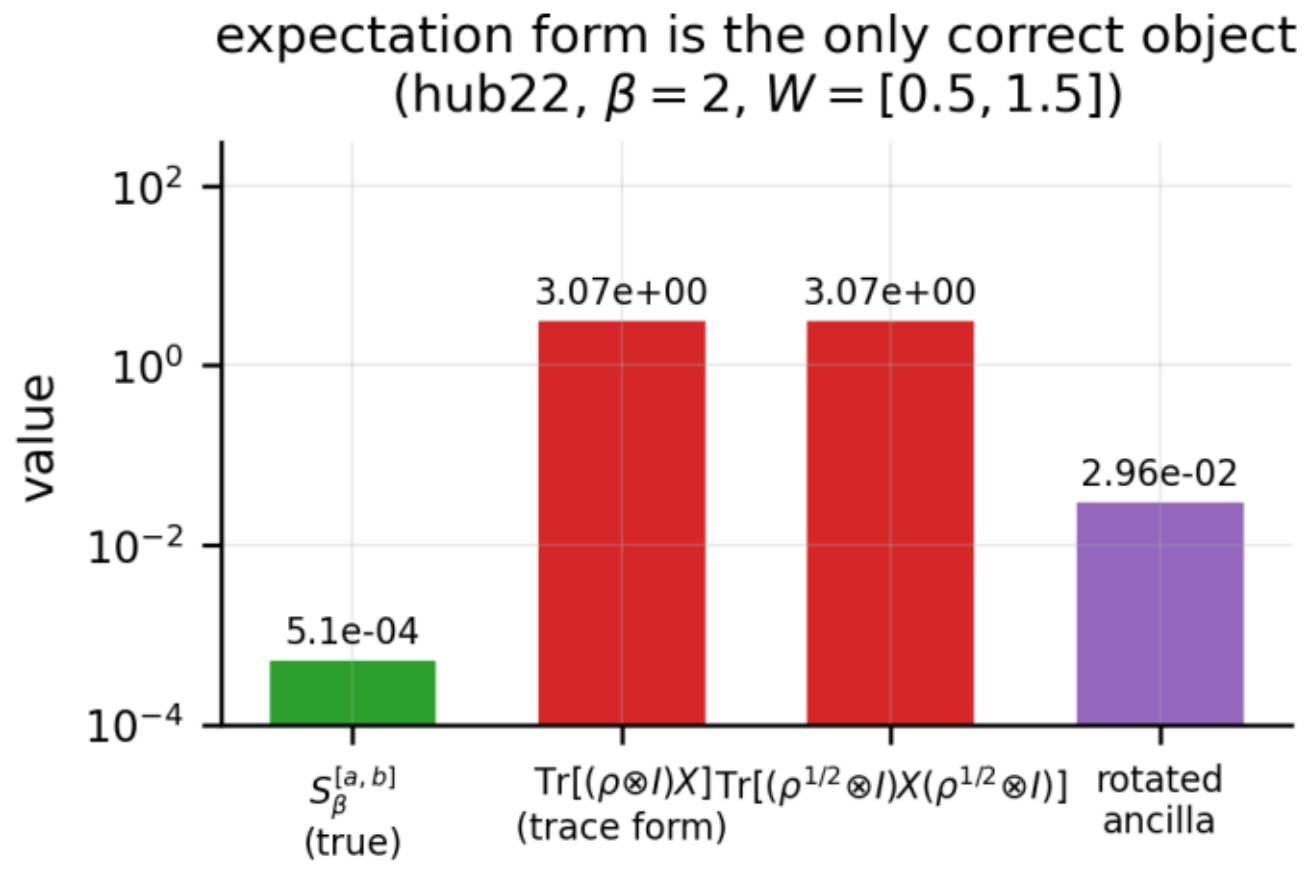

*Fig. 5. The trace form evaluates a different object, numerically. Hubbard* $2 \times 2, \beta = 2,$ *window [0.5,1.5],* log *scale: the true spectral weight S* = $5.097 \times 10^{-4}$ *against the two cyclic-trace forms (both 3.0738651, identical by trace cyclicity and equal to the directly computed density-of-states object to 15 digits) and the expectation under a non-conjugate rotated ancilla* ($2.956 \times 10^{-2}$). *Only the expectation-value form on the Schmidt/TFD purification computes S.*

Remark 3.2 (purification convention). $\langle\Phi_\beta|X|\Phi_\beta\rangle$ depends on the ancilla structure of the purification (X acts on both copies): replacing the ancilla by $|\psi_j\rangle \otimes U|\psi_j\rangle$ with a non-conjugate unitary U changes the value ($5.097 \times 10^{-4} \to 2.956 \times 10^{-2}$ for a fixed-seed random U; V3b). Purification-independence $\langle\Phi|A \otimes I|\Phi\rangle = \mathrm{Tr}[\rho A]$ holds only for single-copy operators. The Schmidt/TFD convention is mandatory, and (importantly for the algorithm) two preparation routes produce exactly this state: (i) the direct QSP preparation of Lemma 7, $(e^{-\beta H/2} \otimes I)|\Omega_c\rangle$ with $|\Omega_c\rangle$ the computational-basis maximally entangled state, whose energy-basis form $\sum_j |\psi_j\rangle|\psi_j^*\rangle/\sqrt{d}$ is compatible with the energy-diagonal kernel $e^{-\beta\lambda/2}$; and (ii) the coherent algorithm of Chen et al. [7], whose output approximates the canonical purification $\sum_i e^{-\beta E_i/2} |\psi_i\rangle|\psi_i^*\rangle/\sqrt{Z}$ in vector norm (their Thm. I.4, Eq. (49)); the approximation error enters the error budget exactly as a preparation error (App. B). Mixed-state and the tested typicality preparations cannot supply the second-copy coherence for the *duplex* readout (Sec. 5.1, with the scope qualifier there); mixed realizations feed the bin and Fourier-LCU routes.

Remark 3.3 ($D' \neq D$ and tensor components). The same contraction gives, for two Hermitian operators $D' \neq$ D,

$$S(D',D)_\beta^{[a,b]} \equiv \sum_{m,n} p_m\, d'_{mn} d_{nm} \mathbf{1}_{[a,b]}(\lambda_n - \lambda_m) = \langle\Phi_\beta|(D' \otimes I)\mathbf{1}_{[a,b]}(H_\Delta)(D \otimes I)|\Phi_\beta\rangle, (T3')$$

because the bra-side $(D' \otimes I)$ contributes $\langle j|D'|n\rangle$ against the ket-side $\langle n|D|j\rangle$ (the matrix product of the two energy-basis matrices enters transposed; App. A.4). Tensor components $\chi_{ij}$ (response along i to a perturbation along j: $D' = D_i$ on the bra side, D = $D_j$ on the ket side) are the response-theoretic instance. Numerically (P5, Hei8, D = $S^z_{q=\pi}, D' = S^z_{site0}$; $\beta \in \{0.5,2,8\}$, two windows): construct vs. SOS agree to max relative $3.1 \times 10^{-16}$. The Hadamard test reads Re of the amplitude, as in the zero-temperature $D'$ ≠ D case; the diagonal case $D' = D$ of Theorem 3 is real non-negative.

Circuit structure and cost (component-wise). $H_\Delta = H_0 \otimes I - I \otimes H_0^T$ is a two-term LCU of block-encodings, realized by a routed SELECT: the select qubit is prepared in $(|0\rangle + |1\rangle)/\sqrt{2}$, and SELECT = $|0\rangle\langle 0|\otimes U_H^{(\text{copy }1)} - |1\rangle\langle 1|\otimes U_H^{T(\text{copy }2)}$ (the sign on the second branch is essential; without it the sandwich computes the sum operator) is implemented by two layers of controlled-SWAPs over the two copies' system Hamiltonian registers (log d qubits each; the m-qubit primitive ancillas of $U_H$ sit at the primitive slot and are shared/bystander) around one application of the Hamiltonian block-encoding primitive, routed to whichever copy the select qubit addresses, followed by a phase Z on the select qubit that implements the branch sign. The two branches are mutually exclusive, so each $H_\Delta$ query costs one query to the H-family primitive, with a branch-implementation caveat at complex $H_0$: for real $H_0$ the two branch circuits coincide ($U_{H^T} = U_H$), and a single physical primitive is routed to whichever copy the select qubit addresses. For complex $H_0$ the copy-2 branch requires the conjugate circuit $U_H^*$ (synthesized at identical LCU cost by conjugating the coefficients), which is a *different* circuit from $U_H$; the SELECT then implements a controlled multiplex between the two; the query count per $H_\Delta$ query remains one, at an additional controlled-switch cost of O(m) ancillas/gates (Pauli-LCU realization: controlled sign flips on the Y-term coefficients): the $H_\Delta$ block encoding is $(2\alpha, m + 1, 0)$, for real $H_0$ and $(2\alpha, m + 1 + O(m), 0)$, in the complex case. In either case the QSP degree for a width-$\gamma_w$ window is $\tilde{O}((2\alpha^{\text{eff}}/\gamma_w) \cdot \log^2(1/\varepsilon'))$ *in $U_H$-family queries*; $d_W$ counts $H_\Delta$ queries and $U_H$-family queries identically. The window-width $\gamma_w$ now means *transition-frequency resolution* (the spectrum of $H_\Delta$ is exactly the transition-frequency set; verified numerically to $5 \times 10^{-14}$, V2). The dipole slots act on the first copy and reuse the zero-temperature components; the readout coin is the all-flag post-selection probability of App. C with $\zeta_\beta = 2\alpha \cdot \alpha_D^2$ under the safe convention. Everything except the initial state and the filter target is the zero-temperature machine, Table 2. (Total per-window costs are multiplicative in the preparation; Sec. 6.)

Table 2. Component-wise comparison: zero-temperature vs. thermal duplex readout (structural; per-window query formulas from Theorem 10 / the readout theorem of App. C; both machines assembled with the same exact-reflection convention).

| component | zero-T [1] | thermal duplex (this paper) | change |
|---|---|---|---|
| initial state | $\|\psi_0\rangle$ (eigenstate filtering, A1: overlap $p_0$, gap G) | $\|\Phi_\beta\rangle$ (Schmidt/TFD; Sec. 5 routes) | A1 replaced by explicit preparation parameters (which grow at low temperature) |
| filter operator | $\mathbf{1}_{[a,b]}(H_0)$, $(\alpha, m, 0)$-encoding | $\mathbf{1}_{[a,b]}(H_\Delta)$, $(2\alpha$, m+1, 0)-encoding | window semantics: absolute → transition, routed LCU: 1 H-family query per filter query |
| window width $\gamma_w$ means | absolute-energy resolution | transition-frequency resolution | decoupled from $\lambda$ at $\beta > 0$ |
| dipole slots | $U_D, U_{D'}$ | $U_D \otimes I, U_{D' \otimes I}$ | unchanged (first copy) |

| component | zero-T [1] | thermal duplex (this paper) | change |
|---|---|---|---|
| composite subnormalization | $\zeta_\omega = (\alpha + 1 + \omega)\alpha_D^2$ (safe) | $\zeta_\beta = 2\alpha\alpha_D^2$ (safe) | safe-convention ratio $2\alpha/(\alpha + 1 + \omega)$, tight normalization: ratio 1 |
| readout per window (filter + AE) | $M_{\rm AE}^0 \cdot 4 \cdot d_W^0$ per iterate, $d_W^0 = O(((\alpha + \lVert c\rVert)/\gamma_w) \cdot \log^2(2/\varepsilon'))$ under the paired shift | $M_{\rm AE} \cdot 4 \cdot d_W$ per iterate, $d_W = O((2\alpha^{\rm eff}/\gamma_w) \cdot \log^2(2/\varepsilon'))$ | readout-stage ratio $(2\alpha + \lvert c\rvert)/(\alpha + \lvert c\rvert) \approx 1.79$ with paired shift conventions (Cor. 10.1); readout stage only — the full per-window total (preparation included) is Eq. 10-sum |
| free self-checks | — | FDT-L mirror + weight-level detailed balance | 0 extra queries when mirrored windows are already measured |

## 4.2 Single-copy bin decomposition (route D): asymmetric widening within the spectral-filter class

Proposition 4. Partition the $\lambda$ axis into bins $I_k = [u_k, u_k + \Delta)$ of width $\Delta$ covering the thermally occupied range $[0, E_{\rm cut}]$, and set $I_k + W = [u_k + a, u_k + \Delta + b]$. Then

$\sum_k \mathrm{Tr}\,[\rho_\beta \cdot \mathbf{1}_{I_k}(H_0) \cdot D\mathbf{1}_{I_k+W}(H_0)D] = \sum_{m,n} p_m\, \mathbf{1}_{\lambda_m \le E_{\rm cut}}\, |d_{mn}|^2 \mathbf{1}_{[a-t_m, b+\Delta-t_m]}(\lambda_n - \lambda_m)$, (P4)

with $t_m = \lambda_m - u_{k(m)} \in [0, \Delta)$ the offset of the initial state inside its bin; the widening is per-initial-state *asymmetric* (left edge shrunk by $t_m$, right edge extended by $\Delta - t_m$); the total is contained in the symmetric envelope $[a - \Delta, b + \Delta]$. Each summand is an $Id - 1 - \beta$ double-slot instance (circuit-wise the nested (Eq. 27)-type product with a projector slot) so the route reuses higher-order machinery on a single copy.

*Proof.* Expand by $Id - 1 - \beta$; $\lambda_m \in I_k$ and $\lambda_n \in I_k + W \Leftrightarrow \lambda_n - \lambda_m \in [a - t_m, b + \Delta - t_m]$ with $t_m = \lambda_m - u_k$. (Full proof: App. A.5.)

Corollary 4.1 (separation rank; remark-level, two layers). *Layer 1 (continuous kernel, uniform bound).* For the exact difference-window indicator as a kernel on $\mathbb{R}^2$, there is no spectrum-independent finite separation rank: for any K, an explicit deterministic pair of grids (App. A.5) makes the $K' \times K'$ matrix $\mathbf{1}_W(x_i - y_j)$ unit-banded (bandwidth 1 < $K'$ = K+1) and full rank with determinant exactly 1, covering all window types (positive-frequency, zero-straddling, and degenerate point windows); so any separated form $\mathbf{1}_W(x - y) = \sum_{k \le K} f_k(x) g_k(y)$ valid on *all* of $\mathbb{R}^2$ needs ≥ K+1 terms, for every K.

*Layer 2 (fixed finite spectrum).* On a fixed finite-dimensional system the kernel restricts to the finite matrix $M_{mn} = \mathbf{1}_W(\lambda_n - \lambda_m)$, which *necessarily has finite rank* ≤ d; a finite separated representation on those spectrum points always exists (e.g. the eigen-decomposition of M), and nothing in Layer 1 forbids it. What Layer 1 does constrain is *uniformity and cost*: any such representation adapted to a fixed spectrum may use spectrum-dependent factors, and its LCU coefficient norm can grow with system size and with the spectrum's fine structure; a representation with spectrum-*independent* factors and bounded term count would specialize to the grids of Layer 1 and cannot exist. *The spectral-filter class, defined.* By "absolute-energy spectral-filter circuit" we mean: one copy, an initial state (possibly with ancillas) diagonal in the energy basis, and a circuit whose only $H_0$-dependence is through spectral projectors/functions $f_k(H_0)$ inserted between the dipoles; so that the measured object is $\mathrm{Tr}[\rho_\beta \sum_k$ (products of $f_k(H_0)$ and dipoles)] whose SOS expansion factors per copy (finite-rank separated kernel with the $f_k$ as factors); ancilla registers are allowed but the kernel factors must remain functions of one copy's spectrum, and no spectrum-adapted (energy-data-dependent) factors are permitted. Within this class at a *fixed bin budget*, the practical content is Prop. 4's per-initial-state asymmetric widening $[a - t_m, b + \Delta - t_m]$; a property of the chosen finite bin realization, not a universal obstruction to single-copy quantum computation: when spectral lines are separated by ≥ 2Δ and bin edges avoid the lines, the widening does not merge lines (the ramp condition of App. B), and the single-copy *time-domain* route lies outside the spectral-filter class (its kernel $e^{-i(x-y)t}$ = $e^{-ixt} \cdot e^{+iyt}$ is rank-1 separated and carries exactly the transition spectrum (developed as the fourth readout route, Fourier-LCU, in Sec. 5.1). What distinguishes the routes is the cost structure of the separation (the LCU coefficient norm and the achievable window sharpness per query), not feasibility: every *approximate* window (on one copy or on two) is a finite-degree polynomial, and any polynomial $P(x - y)$ of degree d is a separated kernel of rank ≤ d+1 (the duplex route's own realization included). Bin count = $E_{\text{cut}}/\Delta$ with thermal support $E_{\text{cut}} = O(\beta^{-1}\ln(d/\delta))$; the bin count reaches 1 (recovering the zero-temperature circuit, a structural consistency check) only once $\beta \gtrsim \ln(d/\delta)/\Delta$.

The symmetric form $[a - \Delta, b + \Delta]$ is only the envelope, the exact widening is asymmetric (V5: the bin sum equals the asymmetric rearrangement to rel = 0.0, a pure rearrangement; the symmetric form differs by 4.5%).

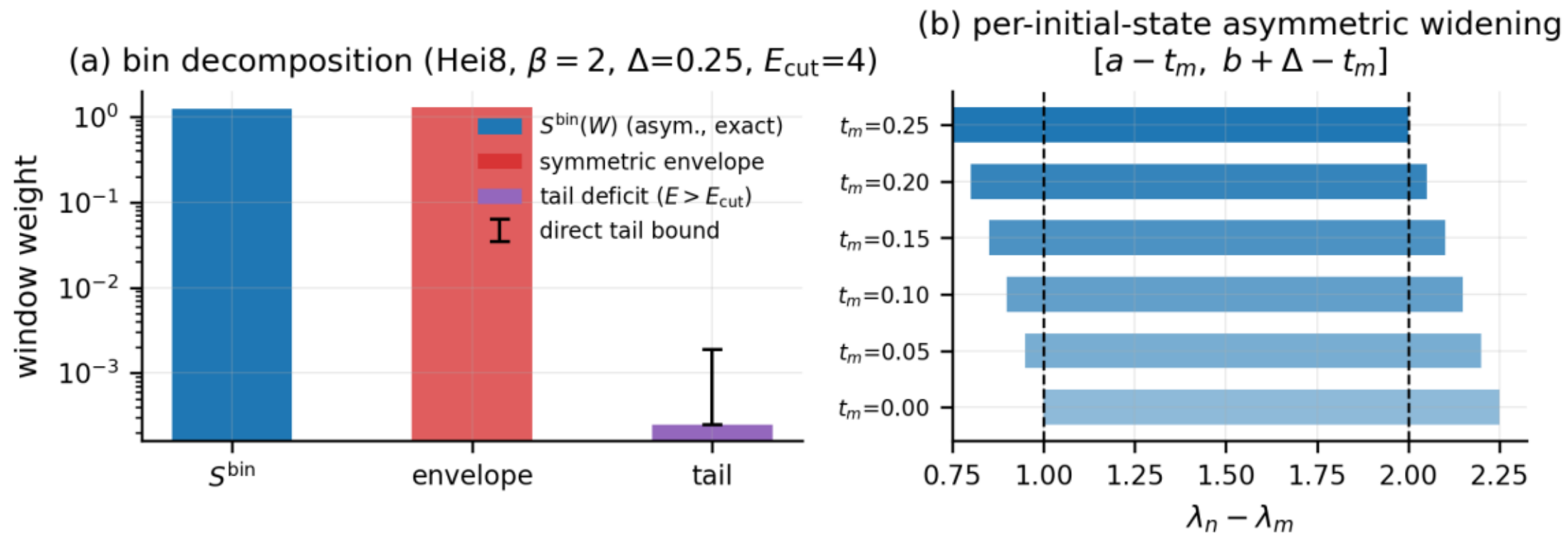

*Fig. 6. Bin decomposition and its asymmetric widening. (a) Heisenberg ring, $\beta = 2$, window W = [1,2], bin width $\Delta = 0.25$, thermal cutoff $E_{cut} = 4.0$ (24 initial states above cutoff): the bin sum $S^{bin} = 1.2295$ (blue; equals the asymmetric-widening rearrangement to rel = 0.0), the symmetric envelope value 1.2877 ≥ $S^{bin}$ (orange), and the tail deficit $2.48 \times 10^{-4}$ ≤ direct bound $1.88 \times 10^{-3}$ (violet with bound whisker). (b) The widened window $[a - t_m, b + \Delta - t_m]$ as a function of the in-bin offset $t_m$ ∈ [0, 0.25]: left edge shrinks, right edge extends, asymmetric per initial state.*

## 4.3 Single-sided filtering is information-insufficient

Proposition 5. The single-sided data J(W) = $\text{Tr}[\rho_\beta D \mathbf{1}_W(H_0) D]$ = $\sum_n \mathbf{1}_W(\lambda_n) c_n$ with $c_n = \sum_m p_m |d_{mn}|^2$ (the thermal column weights; all windows' J-values are equivalent to the sequence $c_n$) do not determine $S_\beta$. *Counterexample (four levels, two channels, V6)*: $\lambda$ = (0, 1, 1.5, 3), $\beta = 1$. System A: channels (0 → 1) with strength S and (2 → 3) with strength T = $p_0 p_1 S/(p_2 p_3)$, system B: channels (0 → 2) with strength $S' = p_1 S/p_2$ and (1 → 3) with strength $T' = p_0 S/p_3$. Then $c^A = c^B$ identically (numerically to $8.9 \times 10^{-16}$) while $S_A$ has lines at $\omega = 1$ and 1.5 and $S_B$ at $\omega = 1.5$ and 2.0: $S_A^{[0.9,1.1]} = 0.6095 \neq 0 = S_B^{[0.9,1.1]}$ (Fig. 7). A three-level fully-connected attempt fails (the matching constraints force the channel strengths to coincide), which is why the counterexample needs four levels and two channels. (Full proof: App. A.6.)

The scope of the counterexample: it equates the two systems on *ideal sharp* windows (all J-values ⟺ the column-weight sequence $c_n$); practical QSP windows have finite width and edge error, smoothing the equivalence class accordingly. The implication for algorithm design: the thermal window weight needs the *joint* (initial, final) distribution, not a marginal; implementing it exactly is precisely what Theorem 3 (window hangs on $\lambda_n - \lambda_m$ automatically) and Prop. 4 (bins + widening) do.

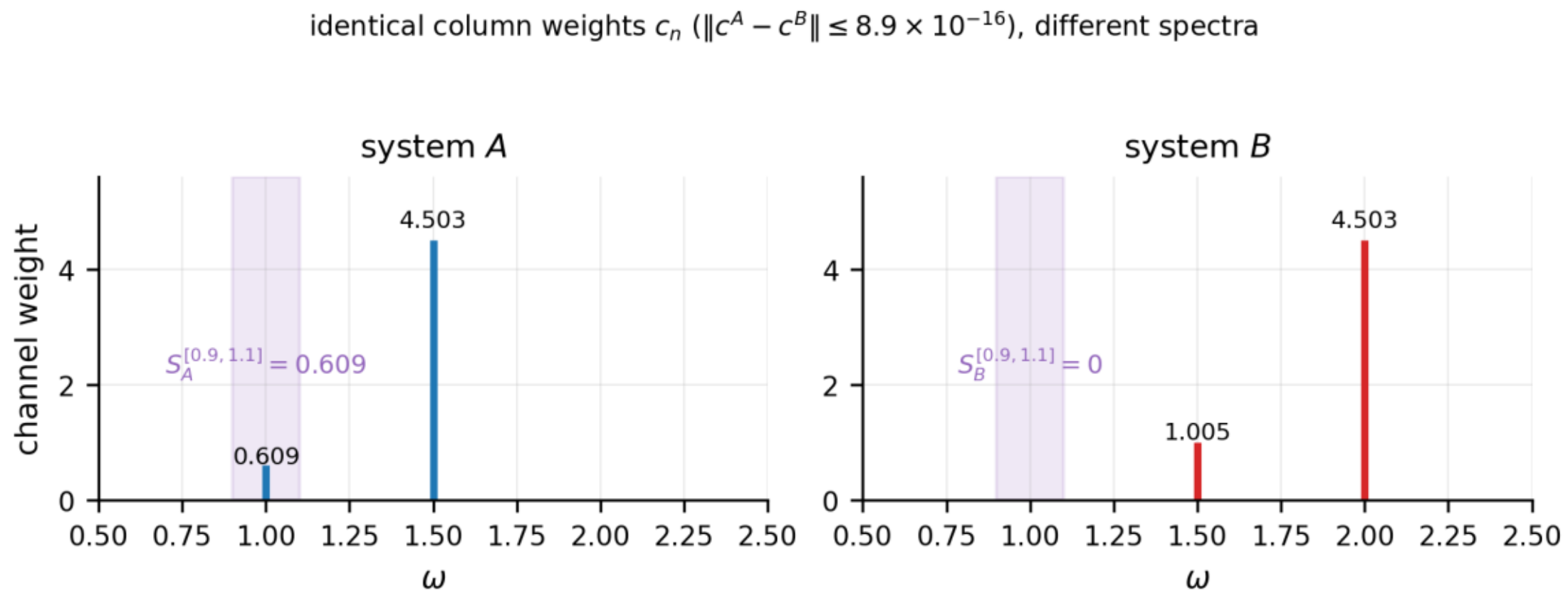

*Fig. 7. Four-level counterexample to single-sided filtering. λ = (0,1,1.5,3),* $\beta = 1$. *System A (blue) and system B (red) have identical column-weight sequences* ($c^A = c^B$ *to* $8.9 \times 10^{-16}$) *but different spectra; the shaded band [0.9,1.1] separates them* ($S_A = 0.6095 vs. S_B = 0$)*. Line labels are the channel weights* $p_0 S$ *etc.*

## 4.4 Higher order: the chain of transition frequencies

Observation (higher-order denominators). Each SOS permutation term of the thermal n-th-order response carries the denominator chain $\lambda_{j_k} - \lambda_{j_{k-1}} - \omega_k - i\gamma^{-1}$, $k$ = 1..n, n transition frequencies referencing n+1 energies. In the Liouville-space picture (Sec. 1.3; the standard framework of nonlinear spectroscopy [2]), the n-th-order thermal response is the double-sided Feynman-diagram object $\langle\langle D|\mathcal{G}(\Omega_n)\mathcal{D}^\times \cdots \mathcal{G}(\Omega_1)\mathcal{D}^\times|\rho_\beta\rangle\rangle$, where $\mathcal{G}(\Omega) = (\Omega - \mathcal{L}_H + i\gamma)^{-1}$ is a function of the same commutator superoperator $\mathcal{L}_H = [H,\cdot]$; whose matrix representation is exactly the doubled register $H_\Delta$ that Theorem 3 block-encodes. Each intermediate coherence $|a\rangle\langle b|$ sits at the $H_\Delta$ eigenvalue $\lambda_a - \lambda_b$, so repeatedly applying functions of the same $H_\Delta$ register supplies every difference in the chain, in exact structural parallel with the zero-temperature nested circuit reusing $f_k(H_0)$ on a single register. No additional copies and no multivariate QSP are needed in principle; the input $vec(\rho_\beta)$ is preparable by the route-(i) machinery of Sec. 5.2. (Cost caveat: each resolvent $\mathcal{G}(\Omega_k)$ is a spectral function of $H_\Delta$ with block-encoding subnormalization growing with the frequency offset, $O(2\alpha + |\Omega_k| + \gamma)$-scale; the higher-order query cost grows with the largest offset in the chain, so the same-register statement is structural, not a constant-cost claim.) We do not develop the higher-order construction here: the per-difference filter widths, the double-sided diagram bookkeeping, and the resource accounting are future work; this subsection records the observation, in the Liouville formulation the duplex register's single difference slot is *not* an obstruction to n ≥ 2; the chain lives on the same register. (Full proof: Sec. A.7.)

## 4.5 Zero-temperature limit and the ground-state criterion

Cor. 6. As $\beta \to \infty (\lambda_0 = 0$, non-degenerate ground state): $p_m \to \delta_{m0}; Id - n - \beta \to$ the zero-temperature nested identity; $S^L_\beta \to$ the ground-state Lorentzian spectrum; (FDT-L) $\to \mathrm{Im}\alpha^{(1)} = L(\omega) - L(-\omega)$ (base-paper Eq. (54), term-by-term); Theorem 3's duplex window $\to \langle \psi_0 | D 1_{[} a, b](H_0) D | \psi_0 \rangle$ (the base-paper Eq. (5)); the bin count of Prop. 4 → 1 and the widening → the identity. Deviation bound:

$|S^W_\beta - S^W_{T=0}| \le (1 - p_0) \cdot \xi^2 + \sum_{m \ge 1} p_m \cdot \langle m | D^2 | m \rangle \le (1 - p_0)(\xi^2 + \|D\|^2) \le 2(d - 1)\|D\|^2 \cdot e^{-\beta G}$,

with $\xi^2 = \sum_n |d_{n0}|^2 = \langle 0 | D^2 | 0 \rangle$, $\|D\|$ the operator norm, and G the spectral gap (the first inequality is the sharp two-term form: the ground-state column shrinks by $1 - p_0$ while each excited initial state m contributes at most $p_m \|D|m\rangle\|^2 \le p_m \|D\|^2$; the last step uses $1 - p_0 = (Z - 1)/Z \le (d - 1)e^{-\beta G}$). Here $\varepsilon$ is the absolute error of the *window weight* ($|S^W_\beta - S^W_{T=0}|$, Eq. 2, not a pointwise response error). This gives the ground-state-approximation criterion as a sufficient condition

$\beta G \ge \ln(2(d - 1)\|D\|^2 / \varepsilon) \implies$ the zero-temperature circuit outputs the thermal answer to accuracy $\varepsilon$. (Cor. 6)

*(Degenerate ground states and the $\xi^2$-shortcut caveat: App. H; the bound above is stated for a non-degenerate ground state, and the correct bound carries $\|D\|^2$ rather than $\xi^2$.)*

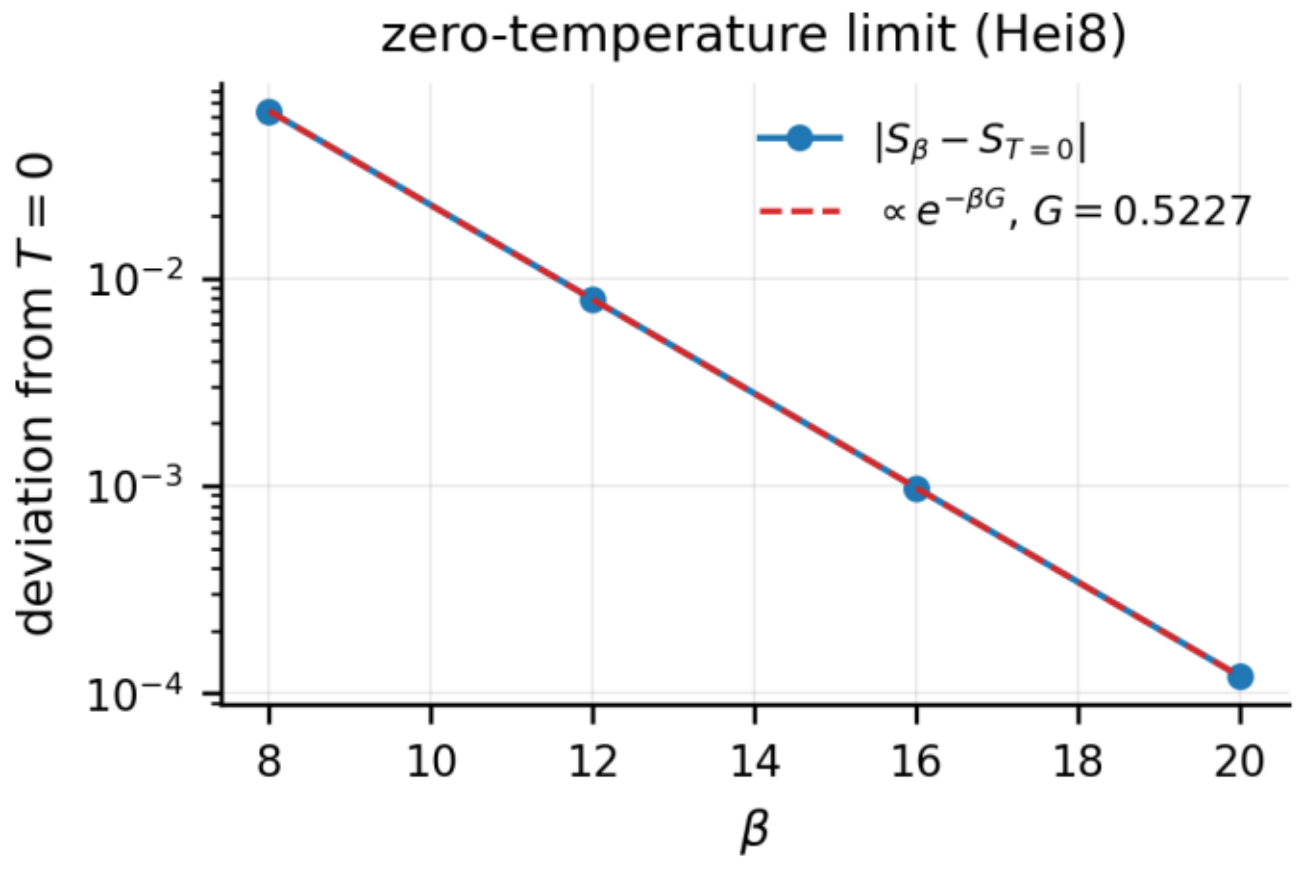

*Fig. 8. Zero-temperature limit. Heisenberg ring N = 8: $|S_\beta - S_{T=0}|$ (window [1,2]) vs. $\beta$ on semilog axes, with the $e^{-\beta G}$ slope (G = 0.52267 from the same spectrum). Fit slope* $-0.52262$ *(0.01% from* $-G$*); adjacent ratios match* $e^{-4G} = 0.12360$ *to < 0.2%; the* $\beta = 20$ *deviation* $1.2099 \times 10^{-4}$ *≤ direct bound* $1.2100 \times 10^{-4}$ *and ≤ sharp Cor. 6 bound* $6.60 \times 10^{-4}$ *(T5a).*

---

## 5. Preparation routes and the decision table

The thermal algorithm's total cost is (preparation) × (readout structure), and the two are not independently choosable: the readout object dictates what the preparation must deliver. Moreover, under the multiplicative accounting of Theorem 10, the preparation cost enters every coherent readout iteration, so preparation-route comparisons must be made per-iterate, not as one-time overheads.

## 5.1 Compatibility matrix

Route F (single-copy Fourier-LCU; fourth readout route). The transition-window weight is also accessible *without any purification*, through the time-domain correlation on one copy. Let g be a smoothing of the window indicator $\mathbf{1}_W$ (e.g. a Gaussian- or erf-smoothed indicator of width $\gamma_w$) with Fourier coefficients $c_k$ over a time grid $t_k$ spanning up to the resolution time; then

$S_\beta^W \approx \sum_k c_k \cdot \mathrm{Tr}[\rho_\beta e^{iHt_k} D\ e^{-iHt_k} D] = \mathrm{Tr}[\rho_\beta \cdot g(ad_H)(D) \cdot D]$,

a single-trace expectation of one-copy operators ($ad_H$ the commutator superoperator; $e^{iHt} D\ e^{-iHt}$ its propagator). The route's target object is the smoothed-window weight $S_\beta(g) = \sum_{m,n} p_m\, |d_{mn}|^2 g(\omega_{mn})$, not the sharp indicator weight $S_\beta^W$ (they differ by $\sum p|d|^2 \cdot |g - \mathbf{1}_W|$, dominated by the line weight within the smoothing band at the window edges: the same edge-sensitivity as the ramp term, now declared as the target's own definition rather than an error). With g an erf-smoothing of the indicator at width $\gamma_w$, the three error/cost knobs are:

- Fourier truncation: the smoothing g of width $\gamma_w$ has Fourier coefficients $c_k$ decaying on the time scale $t_{\max} = O((1/\gamma_w) \cdot \log(1/\varepsilon_F))$ (frequency units for $\gamma_w$, time units for $t_{\max}$; the product $\gamma_w \cdot t_{\max}$ is the dimensionless resolution parameter), with $\|c\|_1 = O(\log(1/\varepsilon_F))$; the quadrature error of the smoothed target is $\leq \varepsilon_F \cdot \mathrm{Tr}[\rho_\beta D^2]$;
- Hamiltonian simulation: each summand is an LCU of two evolutions $e^{\pm iHt_k}$, $|t_k| \leq t_{\max}$, to precision $\varepsilon_{\text{sim}}$ at query cost $O(\alpha \cdot |t_k| + \log(1/\varepsilon_{\text{sim}}))$ per evolution; total simulation cost $\tilde{O}(\|c\|_1 \cdot \alpha \cdot t_{\max}) = \tilde{O}(\alpha/\gamma_w \cdot \text{polylog})$;
- Readout: Hadamard-test/AE on the single-copy correlator; the correlator is estimated through the dipole block encoding (subnormalization $\alpha_D$, since D is not unitary; $\alpha_D$ enters the AE count exactly as in Theorem C.1, $M_{\text{AE}} = O(\max(\sqrt{(1-p)/p}/\varepsilon_{\text{rel}}, 1/\sqrt{\varepsilon_{\text{rel}}}) \cdot \log)$ with $p = S_\beta(g)/\alpha_D^2$), to relative accuracy $\varepsilon_{\text{rel}}$.

Joint bound (Õ-level remark): the total per-window cost of estimating $S_\beta(g)$ to relative accuracy $\varepsilon_{\text{rel}} + O(\varepsilon_F \cdot \text{Tr}[\rho_\beta D^2]/S_\beta(g)) + O(\|c\|_1 \cdot \varepsilon_{\text{sim}} \cdot \text{Tr}[\rho_\beta D^2]/S_\beta(g))$ (the truncation and simulation errors normalized to the target, as in App. B's treatment of $\varepsilon'$) is of order Õ( $\alpha/\gamma_w$ ) $U_H$-family queries per $\rho_\beta$ realization, the same order as the duplex route's window degree, with the smoothing-width $\gamma_w$ playing the role of the resolution parameter; the constants ($\|c\|_1$ and the simulation logarithms) are not developed to the same level as Theorem 10's accounting. Any $\rho_\beta$ realization feeds this route (mixed Gibbs states, typicality draws, METTS trajectories) because the object is a single-copy expectation; this is the natural single-trace form (Remark 3.1). The route's trade-offs relative to duplex: the target is the smoothed weight (with the edge-band weight *in* the target rather than an error term), the readout is a correlator estimation rather than the one-step duplex projection, and the comparison of practical constants is system-dependent (not developed further here).

| readout object | required initial state | mixed $\rho_\beta$ usable? |
|---|---|---|
| Δ (Theorem 3): S = $\langle\Phi_\beta\|(D\otimes I)\mathbf{1}_W(H_\Delta)(D\otimes I)\|\Phi_\beta\rangle$ | Schmidt/TFD canonical purification (ancilla = energy-basis conjugate, fully correlated coefficients), to preparation error within the budget | ✗ for the two typicality constructions tested (independent pairs / same-state square; below) — other constructions not excluded |
| bin (Prop. 4): $\sum_k \text{Tr}\,[\rho_\beta \mathbf{1}_{I_k} D \mathbf{1}_{I_k+W}$ | any $\rho_\beta$ realization (mixed / purified / typicality / METTS) | ✓ |
| F (route above): $\sum_k c_k \text{Tr}[\rho_\beta e^{iHt_k} D$ $e^{-iHt_k}$ D] | any $\rho_\beta$ realization | ✓ |
| adaptive-bin per trajectory (METTS, Sec. 5.4) | METTS trajectory states | ✓ (adaptive) |

Which preparation routes produce the required purification. (i) Direct QSP preparation (Lemma 7): produces $|\Phi_\beta\rangle = (e^{-\beta H/2}\otimes I)|\Omega_c\rangle/\sqrt{Z/d}$ exactly, to trace distance $\varepsilon_p$. (ii) The coherent thermal-state algorithm of Chen et al. [7]: their Theorem I.4 (Eqs. (48)–(49)) outputs a state within vector-norm error $\tilde{O}(\beta/(\sigma_t\lambda_{gap}(\boldsymbol{\mathcal{D}}_\beta)))$ of the canonical purification $\sum_i e^{-\beta E_i/2}\,|\psi_i\rangle|\psi_i^*\rangle/\sqrt{Z}$, the exact Schmidt/TFD structure; the error propagates to the window weight as $|\Delta S| \le 2\varepsilon\|X_W\|$ and enters the error budget of App. B like any preparation error. (iii) Mixing-based routes (Metropolis [11,13], Lindbladian [14,15], Kuwahara's classical quasi-linear algorithm [16]) output (or sample) the *mixed* Gibbs state or a classical description of it: Δ-incompatible *without additional processing* (a classical MPO description can in principle be converted to a $\rho^{1/2}$ purification preparation at additional cost, route-dependent, not analyzed here); bin- and F-compatible directly. Typicality (route ii of Sec. 5.3): bin and F routes directly; for the duplex readout, the two constructions tested fail (below); other typicality constructions are not excluded by our analysis.

Why the two tested typicality constructions do not feed the duplex readout (analytic + numeric; limited to those constructions). The duplex object requires the *complete correlation* of the two copies' energy coefficients (ancilla pinned to the same level j). The deployed two-copy typicality construction (both copies filtered and normalized) has average $\sum_{j,n}\rho\,(j,n)\mu_j\mu_n$ with $\rho(j,n) = \sum_k |d_{kj}|^2\mathbf{1}_W(\lambda_k-\lambda_n)$ and $\mu_j = E|c_j|^2 \neq p_j$: the normalization factor D = $\sum_l e^{-\beta\lambda_l}\,|x_l|^2$ fluctuates and couples to $|x_j|^2$ (E[1/D] > 1/E[D] by a Jensen-type argument, with level-dependent direction). Measured (Hei8, M = 400):

$\mu_0 = 0.4012$ vs. $p_0 = 0.4939$ at $\beta = 2(-18.8$; total-variation distance of $\mu$ vs. p: 0.019 /0.093/0.044 at $\beta$ = 0.5/2/8 (P4, Fig. 10b); both typicalized duplex constructions miss the spectral weight by ~20% at M = 400 with no convergence trend (W1b: 20.5% /19.6%; Fig. 10a); same-state-squared fourth-moment terms contribute a $\mu_j^2$-type *inelastic* sum, also ≠ S.

Scope of this statement. The impossibility evidence covers the two deployed constructions (independent random pairs; same-state square) under per-state normalization without cross-copy correlation. It does *not* rule out other typicality-based routes to the duplex object: a construction proposed as an alternative, random-phase product states $|\phi\rangle \otimes |\phi^*\rangle$ (each copy the conjugate of the same random state), whose second moment contains a $|\Omega_c\rangle\langle\Omega_c|$ component, is claimed to recover S unbiasedly after subtracting separately estimable terms (reported Monte Carlo 1.03 ± 0.07 against the true 1.081 on Hei8 at $\beta$ = 2). Two evaluations of the raw two-copy expectation disagree at the operator/input convention level ($\approx 950\sigma$ apart); we report both without adopting either, with the archived values in App. F. More generally, typicality estimators targeting $vec(\rho^{1/2})$ directly, or correlated-product purifications with normalization corrections, are outside the class tested here. Route (i) itself is the typicality limit of this idea (the maximally entangled $|\Omega_c\rangle$ as the deterministic "typical" input).

## 5.2 Route (i): QSP direct preparation (Lemma 7)

Lemma 7 *(direct QSP preparation; the construction of Lemma 4 of Rall [5] in our conventions; the optimal $d_{\mathrm{vt}}$ column below improves the α-scaling of Rall's cost).* Given the $(\alpha, m, 0)$-block-encoding $U_H$, the purification $|\Phi_\beta\rangle = (e^{-\beta H/2} \otimes I)|\Omega_c\rangle/\sqrt{Z/d}$ is preparable to trace distance $\varepsilon_p$ with

$N_{\mathrm{prep,call}}$ = O( $\sqrt{d/Z}$ ) · $d_{\mathrm{vt}} U_H$ queries per preparation call (with Z known or lower-bounded, see the amplification clause below), $d_{\mathrm{vt}}$ = Õ( $\sqrt{\alpha\beta \cdot \log(d/(Z\varepsilon_p))}$ ) (Lemma 7a; optimal form [39])

*Proof sketch.* The unnormalized preparation post-selects with probability Z/d, which sets the $\sqrt{d/Z}$ counting; the polynomial degree comes from approximating $e^{-x}$ on [0, $\alpha\beta$] (optimal approximation [39]), with the raw accuracy tied to $\varepsilon_p$ through Lemma 7a; the unknown-Z branch uses fixed-point amplification [40] with a lower bound $Z_{\mathrm{lb}} \leq Z$. Full derivation, including the evaluation allocation, in App. A.8.

$\beta$-dependence and size-scaling. $\sqrt{d/Z}$ approaches 1 only in the *ordered* limit $\beta \to 0$ taken before N → ∞; at fixed small $\beta$ it is already above 1 (Z(d=70) = 16.14 at $\beta = 0.5$, $\sqrt{d/Z}$ = 2.08). In many-body systems with positive entropy density it grows exponentially at fixed $\beta$. The growth rate per site is $s(\beta) - (-\beta f(\beta))$, with s the sector entropy density; for the Heisenberg Sz = 0 sector $s = \ln 2$, and $\ln 2 - (-\beta f(\beta)) > 0$ at every finite temperature. The rate ln(d/Z)/N is always ≥ 0 (since 1 ≤ Z ≤ d), so $\sqrt{d/Z}$ never drops below 1 and tends to it only in the joint $\beta \to 0$ limit. This is a thermodynamic rate, not a universal law; it is empirically positive on the systems measured.

Measured on Heisenberg rings (N = 8/10/12/14, full spectra): $\sqrt{d/Z} = 2.08 \to 2.53 \to 3.07 \to 3.73$ at $\beta = 0.5$ and $5.88 \to 9.59 \to 15.69 \to 25.74$ at $\beta = 2$. The amplification factor is thus large in typical many-body regimes, the quantitative form of the known thermal-preparation bottleneck [8]; high-temperature relief requires $\beta \to 0$ (Z → d) as an explicit joint limit with N. The governing scale is the per-site rate $s(\beta) - (-\beta f(\beta))$: positive rates give exponential growth, and the fixed-dimension anchors of Table 4 are single-system measurements, not an asymptotic law. At fixed system the $\beta$-trend is as before: $\sqrt{d/Z}$ = 2.08/5.88/8.30 at $\beta$ = 0.5/2/8 on d = 70.

Mixing-based routes replace $\sqrt{d/Z}$ by their own parameters ($t_{mix}$ for Metropolis/Lindbladian [7,11,13], logarithmic-in-n at high temperature [14], system–bath end-to-end [15]); which wins is system-dependent, and the decision table carries both parameterizations. The QSP kernel $e^{-\beta\lambda/2}$ and the filter kernel $\mathbf{1}_{[a,b]}$ are two target functions of the same QSP machine (degrees $\tilde{O}(\sqrt{\alpha\beta \cdot \log(1/\varepsilon_{raw})})$ vs. $\tilde{O}((2\alpha^{\text{eff}}/\gamma_W) \cdot \log^2(1/\varepsilon')))$: the component library $U_H, U_D, QSP$ is fully shared, and no new primitive is needed. Under multiplicative accounting (Theorem 10) this preparation cost is paid per coherent iterate, not once.

## 5.3 Route (ii): thermal typicality (Lemma 8) — allocation-dependent ε-scalings, two estimators, and a per-state preparation cost

Status labels for this subsection (three kinds of content, kept apart). *(T)* exact identities, $E[D_\phi]$ = Z/d, $E[N_\phi - J \cdot D_\phi] = 0$, the Jensen lower bounds, and the second-moment identity for $Var[D_\phi]$: theorem-level. *(O)* measured observations, the ensemble bias b, the variance parameters $\sigma_X/\sigma_b$, the negative-moment ratios, the delta-method heuristics, and the batch-count evidence (finite-sample, anchor-specific). *(S)* protocol sketch, the destructive batched readout of estimator (b) (coarse-grained; all quantitative cost statements default to estimator (a)'s M-formula). Key statements below carry their labels.

Notation (J vs. S). The typicality protocol measures the bin-route absolute-window weight $J_\beta^{[0,3]} = \text{Tr}[\rho_\beta D 1_[0,3](H_0) D]$ (the [0,3] window is the concrete instance used in all our numerics; on Hei8 at $\beta = 2$ it equals 5.7025, whereas the Δ-route transition-window weight $S_\beta^{[1,38]} = 1.0809$, different objects, both quoted henceforth with their window labels).

Lemma 8. With random-phase states $|\phi\rangle = d^{-1/2} \sum_s e^{i\theta_s} |s\rangle$ filtered to $|\psi_\phi\rangle \propto e^{-\beta H/2}|\phi\rangle$, write $N_\phi = \langle\phi|e^{-\beta H/2} D 1_W(H_0) D e^{-\beta H/2}|\phi\rangle, D_\phi = \langle\phi|e^{-\beta H}|\phi\rangle$ (so the per-state estimator is $N_\phi/D_\phi$). Two estimators:

(a) *(per-state ratio mean)* $mean_\phi(N_\phi/D_\phi)$: concentrates around J with typical fluctuation $\sigma_X$ per state *(O)*, and carries a Jensen-type observed ensemble bias b *(O)* ($E[N_\phi/D_\phi] \neq$ J because the normalization couples to the numerator) that more random states cannot remove;

(b) *(ratio of means)* $(\sum_\phi N_\phi)/(\sum_\phi D_\phi) \to$ E[N]/E[D] = J, consistent (asymptotically unbiased): the exact identity is $E[N_\phi - J \cdot D_\phi] = 0$, while the ratio itself has an O(1/M) delta-method bias (App. A.9) plus sampling noise with variance parameter $\sigma_b^2 \equiv Var(N_\phi - J \cdot D_\phi)/E[D_\phi]^2$, so that $RMSE_b(M) \approx \sigma_b/(J\sqrt{M})$ relative to J and $M^{(b)} = O(\sigma_b^2/(\varepsilon_{\text{rel}} J)^2)$.

Per-state preparation costs are governed by negative-moment product expectations, not by $(E[D_\phi])^{-1}$. For a *sampled* state the coherent-amplification cost is $O(D_\phi^{-1/2} \cdot d_{\rm vt}(\phi))$ and the post-selection retry cost is $O(D_\phi^{-1} \cdot \alpha\beta)$, with the state-dependent normalization-amplification degree $d_{\rm vt}(\phi) = O(\alpha\beta + ½\log(1/D_\phi) + \log(1/\varepsilon_p))$ (conservative column; optimal column $\tilde{O}(\sqrt{\alpha\beta \cdot \log(1/\varepsilon_{raw}(\phi))})$) (the per-state form of Lemma 7a: $\varepsilon_{raw}(\phi) = \varepsilon_p \cdot \sqrt{D_\phi}/2$). The primary form of the amplified cost is the expectation of the product, $E[D_\phi^{-1/2} \cdot d_{\rm vt}(\phi)]$ (exact object); the ensemble form $E[D_\phi^{-1/2}] \cdot d_{\rm vt}$($d_{\rm vt}$ evaluated at the ensemble mean; = 21.8/126.6/661.7 at $\beta$ = 0.5/2/8 from the measured $E[D_\phi^{-1/2}]$ = 2.09/6.34/12.20 and $d_{\rm vt}$ = 10.5/20.0/54.2) is an anchor approximation, whose accuracy at our anchors is the measured coupling excess (T5i `d_vt_state_coupling`: 0.03%/0.51%/1.07% at $\beta$ = 0.5/2/8) *(O)*. Averaging over states therefore introduces, by Jensen's inequality,

$$E[D_\phi^{-1/2}] \geq (E[D_\phi])^{-1/2} = \sqrt{d/Z}, E[D_\phi^{-1}] \geq (E[D_\phi])^{-1} = \mathrm{d/Z},$$

with equality only for deterministic $D_\phi$; the identity $E[D_\phi]$ = Z/d licenses only the lower bounds. The fluctuation of $D_\phi$ is exactly characterized (random-phase states, real eigenvectors) by the second-moment identity $Var[D_\phi] = [Z(2\beta) - w^{\rm T}Qw]/d^2 \leq Z(2\beta)/d^2$, with $w_j = e^{-\beta\lambda_j}$ and $Q_{jk} = \sum_s V_{sj}^2 V_{sk}^2$ (App. A.9; $CV^2 = Var/(Z/d)^2$ = 0.0057 /0.2354/0.9028 at $\beta$ = 0.5/2/8 on Hei8), so the negative moments diverge from the Jensen lower bounds exactly as the ensemble widens. Measured on the same 1600-state pools as all typicality experiments (T5i): $E[D_\phi^{-1/2}]$ = 2.08743/6.33656/12.20099 = 1.0025 ×/1.0777 ×/1.4699 × above $\sqrt{d/Z}$, and $E[D_\phi^{-1}]$ = 4.36341/42.29666/228.52947 = 1.0064 ×/1.2235 ×/3.3167 × above d/Z ($\beta$ = 0.5/2/8). The $d_{\rm vt}$ state-dependence couples to the amplification rounds at the ≤ 1.07 level at these anchors in ensemble average (T5i `d_vt_state_coupling`; individual worst states reach ≈ 8 on the amplification-precision subterm; the ensemble average is the cost-relevant quantity); the trivial rigorous upper bound is $E[D_\phi^{-1}] \leq e^{\beta\lambda_{\max}}$, useless at large $\beta$; an effective non-trivial upper bound remains open.

Per-state cost: two error allocations (the ε-scaling is allocation-specific, not intrinsic). With M states and one amplitude estimation per state, the treatment of the per-state AE error determines the total ε-dependence:

- ($\alpha$) systematic-error allocation (each state's AE at absolute accuracy $\Delta = \varepsilon_{\rm rel} \cdot J$, errors added in quadrature as systematic): total $error^2 = \sigma_X^2/M + \Delta^2$, balanced at $\Delta = \varepsilon_{\rm rel} \cdot J$ giving M = $O(\sigma_X^2/(\varepsilon_{\rm rel}J)^2)$ states and per-state $M_{\rm AE}^{\rm bin} = O(\zeta_\omega \cdot \log(1/p_{\rm fail})/(\varepsilon_{\rm rel}J))$ iterates, total

$M \cdot M_{\rm AE}^{\rm bin} \cdot (2 \cdot N_{\rm prep}^{\rm state} + 4 \cdot d_W^{\rm bin}) \Longrightarrow \tilde{O}(\sigma_X^2 \cdot \sqrt{(d/Z)} \cdot d_{\rm vt} \cdot (\alpha + E_{\rm cut})/(\gamma_w \cdot \varepsilon_{\rm rel}^3 \cdot J^3))$-type $U_H$ queries (Lemma 8$\alpha$)

This is the $\varepsilon_{\rm rel}^{-3}$ upper-bound structure (the allocation used in all quantitative statements and tables of this paper);

- ($\beta$) independent-error allocation (each state's AE at a *constant* per-state precision $\delta \sim \sigma_X$, independent across states): the per-state AE errors then average with the state fluctuations, total $error^2 \approx (\sigma_X^2 + \delta^2)/M$, so M = $O((\sigma_X^2 + \delta^2)/(\varepsilon_{\text{rel}} J)^2)$ with $\delta$ ε-independent, and each state's AE runs at constant iterate count $M_{\text{AE}}^{\text{bin}}(\delta) = O(\zeta_\omega \cdot \log(1/p_{\text{fail}})/\delta)$: total ε-dependence $\varepsilon_{\text{rel}}^{-2}$ (or $\varepsilon_{\text{rel}}^{-2.5}$ if the AE bias at fixed count carries a nonlinear-in-p component that must be suppressed as J shrinks, the estimator-(b) sketch of Sec. 5.3 uses exactly this averaging argument, and it applies to estimator (a) alike once the per-state readouts are independent).

The allocation choice is a protocol decision; $\varepsilon_{\text{rel}}^{-3}$ is an upper bound under allocation $(\alpha)$, not an intrinsic scaling of the typicality route. In both variants the per-state preparation is $N_{\text{prep}}^{\text{state}} = O(E[D_\phi^{-1/2} \cdot d_{\text{vt}}(\phi)])$, the expectation of the product (primary form; the ensemble anchor form $E[D_\phi^{-1/2}] \cdot d_{\text{vt}}$ evaluates to 21.8/126.6/661.7 at $\beta$ = 0.5/2/8 from the measured $E[D_\phi^{-1/2}]$ = 2.09/6.34/12.20 and the conservative $d_{\text{vt}}$ = 10.5/20.0/54.2, with coupling excess ≤ 1.07%; the naive $(E[D])^{-1/2}$ convention $\sqrt{d/Z}$ = 2.08/5.88/8.30 is only a Jensen lower bound (Table 3 quotes both); the post-selection variant) $O(E[D_\phi^{-1}] \cdot \alpha\beta)$ queries (measured 4.36/42.30/228.53 vs. the naive d/Z = 4.34/34.57 /68.90), a usable normalized $e^{-\beta H/2}|\phi\rangle$ costing on average $E[D_\phi^{-1}]$ random states since $E[D_\phi]$ = Z/d; requires measurement and applies only to shot-based per-state readout. $d_W^{\text{bin}} = O(((\alpha + E_{\text{cut}})/\gamma_w) \cdot \log^2(2/\varepsilon'))$ is the single-copy bin-window degree with the off-center shift bound ($E_{\text{cut}}$ = the thermal cutoff; applied four times per iterate (conjugate pair in A and $A$ †) as in App. C). *Proof (allocation $\alpha$).* Variance decomposition $\sigma_X^2/M + \Delta^2$ balanced at $\Delta = \varepsilon_{\text{rel}} \cdot J$; per-state AE with $M_{\text{AE}}^{\text{bin}}$ iterates, each carrying two amplified preparations and the conjugate-pair window four times (App. C); per-state preparation costs from the negative moments above (App. A.9). Allocation $(\beta)$ replaces the balancing step as stated. The ensemble bias of estimator (a) is empirical, with a Jensen-type mechanism (App. A.9).

Positioning of the bias b. The bias is a property of *estimator (a) under the per-state-normalization protocol* (each state measured by its own AE, the natural implementation), not an intrinsic limit of thermal typicality: estimator (b) removes it, at the price of batched readout (numerator and denominator accumulated across states). We report both, without a single-sided claim: the per-state protocol has operational value (independent per-state readouts, online stopping) and a measured ensemble bias; the batch protocol is asymptotically unbiased (O(1/M) bias) and needs joint aggregation.

*Protocol detail, the dual-estimator experiment (T5b), the finite-size scope of the measured factors, and the statistical-uncertainty reporting are collected in App. F of the Supplemental Material.*

Applicability precondition for (a): $\varepsilon_{\text{rel}} \gtrsim$ b/J; below the observed bias the estimator is unreachable; for (b) the precondition is the batching overhead itself. The trade (i) ↔ (ii) is $[\sqrt{d/Z} \cdot d_{\text{vt}} periterate]$ vs. $[\sigma_X/(\varepsilon_{\text{rel}} J)]^2$ states + per-state costs: both system-dependent, carried side by side in the decision table; with the multiplicative-accounting caveat that under Theorem 10 the route-(i) preparation is paid $M_{\text{AE}} \approx \zeta_\beta/(\varepsilon_{\text{rel}}\mu_\beta)$ times per window, while route (ii) pays it M times per window set; see Sec. 5.5.

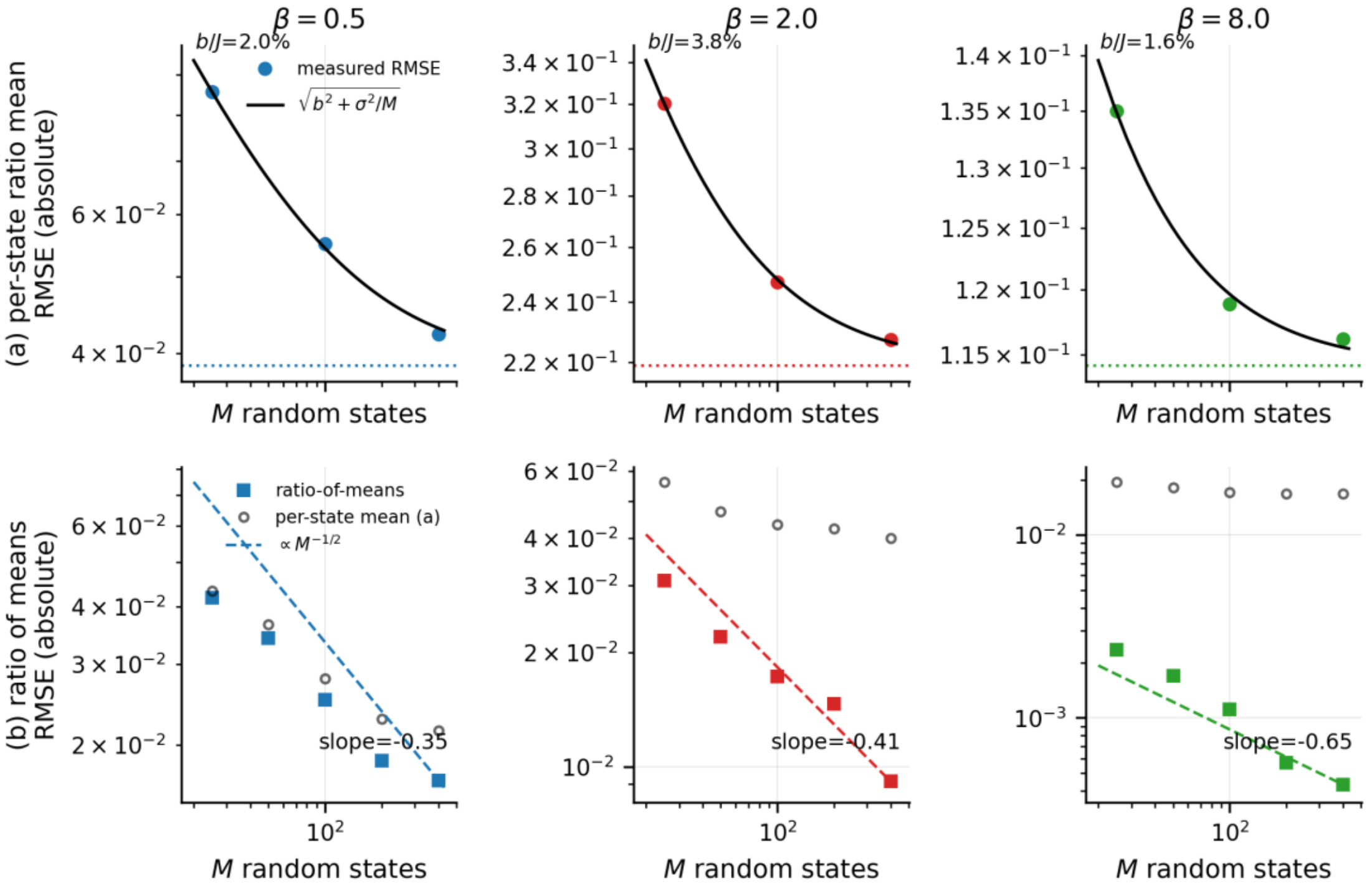

*Fig. 9. Typicality: two estimators. Heisenberg ring N = 8, absolute window [0,3], D = $S^z_{q=\pi}$, target J = $J_\beta^{[0,3]}$; columns $\beta = 0.5,2,8$. Top row (a) per-state ratio mean: RMSE vs. M (log–log), two-component model $\sqrt{b^2 + \sigma^2/M}$; the observed ensemble bias b/J = 1.96%/3.84% /1.65% is flat in M. Bottom row (b) ratio of means: RMSE vs. M decays without a plateau (annotated log–log slopes $-0.35/-0.41/-0.65$ vs. the $1/\sqrt{M}$ reference $-0.5$); open markers repeat estimator (a) for comparison.*

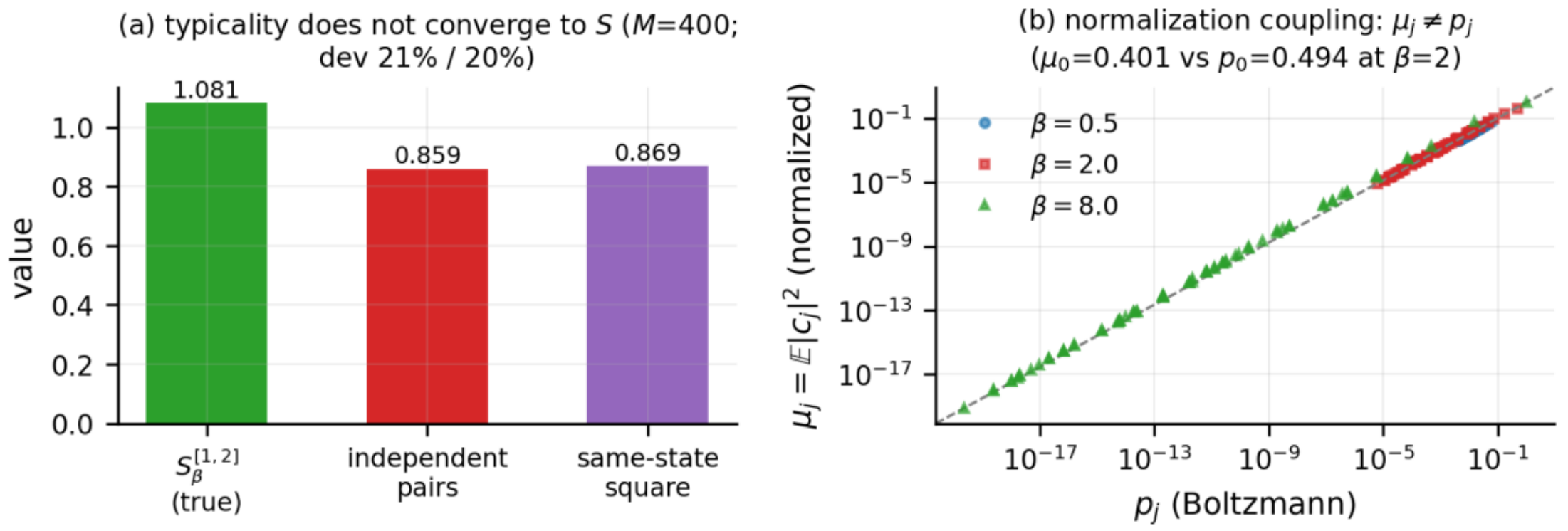

*Fig. 10. Why the two tested typicality constructions do not feed the duplex readout. (a) Duplex expectation under the two typicalized constructions (independent pairs; same-state square) vs. the true transition-window weight $S_\beta^{[1,38]} = 1.081$ at M = 400: deviations 20.5% and 19.6%, no convergence. (b) The normalized energy distribution $\mu_j = E|c_j|^2$ against the Boltzmann $p_j$ (log–log, β = 0.5/2/8): the normalization coupling pushes μ off the diagonal ($\mu_0 = 0.401$ vs. $p_0 = 0.494$ at $\beta = 2$; total-variation 0.093), so the second copy is weighted by μ, not p. (Scope: these two constructions only; other typicality-based routes to the duplex object are not excluded; Sec. 5.1.)*

## 5.4 Route (iii): METTS (Lemma 9) — exact identity and adaptive-bin readout

Route (iii) gives an exact identity and an adaptive-bin readout; its sampling overhead (trajectory count, autocorrelation, burn-in) is characterized empirically, without a complexity theorem.

Lemma 9 (exact identity and adaptive-bin structure). With collapse basis $|c\rangle$ and trajectory states $|\phi(c)\rangle = e^{-\beta H/2}|c\rangle/\sqrt{w_c}, w_c = \langle c|e^{-\beta H}|c\rangle, \pi(c) = w_c/Z$, the METTS identity $\sum_c \pi(c)\langle\phi(c)|A|\phi(c)\rangle = \mathrm{Tr}[\rho_\beta A]$ holds exactly (verified to $2.0 \times 10^{-14}$ over six $(\beta, A)$, combinations, W2a), and the single-trajectory window weight $\langle\phi(c)|D1_{[}a, b](H - \lambda_c^{\mathrm{eff}})D|\phi(c)\rangle$ implements route D with adaptive bins: the trajectory energy center $\lambda_c^{\mathrm{eff}} = \langle\phi(c)|H|\phi(c)\rangle$ is a quantum expectation (one cheap $\langle H\rangle_c$ measurement per trajectory, or $-\partial_\beta \ln w_c$ by adjacent-$\beta$ differencing), after which the window shift costs zero extra QSP degree (target-function translation). *What is rigorous here is the identity and the zero-degree translation.* Three further ingredients are stated with their correct status (all measured, none proved):

*The three measured ingredients, trajectory energy width, entropy participation ratio $e^{S(\pi_\beta)}$, and the low-temperature concentration trend, are collected in App. G of the Supplemental Material; they are measurements on Hei8, not asymptotic claims.*

Total cost structure (sampling overhead open): $N_{total} = \sum_c \hat{\pi}(c) \cdot \sum_W M_{\mathrm{AE}}^c(W) \cdot (2N_{\mathrm{prep,call}}^c + 4 \cdot d_W^{\mathrm{bin}})$ + sampling overhead (open), with $N_{\mathrm{prep,call}}^c = O(\sqrt{(1/w_c)}) \cdot d_{vt}$-type per-trajectory amplified preparation (chain-averaged cost $\le \sqrt{d/Z} \cdot d_{\mathrm{vt}}$ by Cauchy–Schwarz, $E_\pi[\sqrt{1/w_c}] = \sum_c \sqrt{w_c}/Z \le \sqrt{d/Z}$, equality only for uniform collapse weights). Two gaps separate this structure from a complexity theorem: (i) the trajectory count / variance / autocorrelation / burn-in (the sampling-overhead term); (ii) the adaptive bin requires compact per-trajectory energy support, while a trajectory's energy distribution has unbounded tails, centering at $\lambda_c^{\mathrm{eff}}$ controls the mean, not the tails (the measured $\sigma_E(c)$ widths and the heuristic convolution blur above quantify, but do not bound, this effect; a rigorous treatment would need trajectory-energy tail assumptions of the kind Prop. 4's $E_{\mathrm{cut}}$ supplies on the thermal side). (App. A.10.)

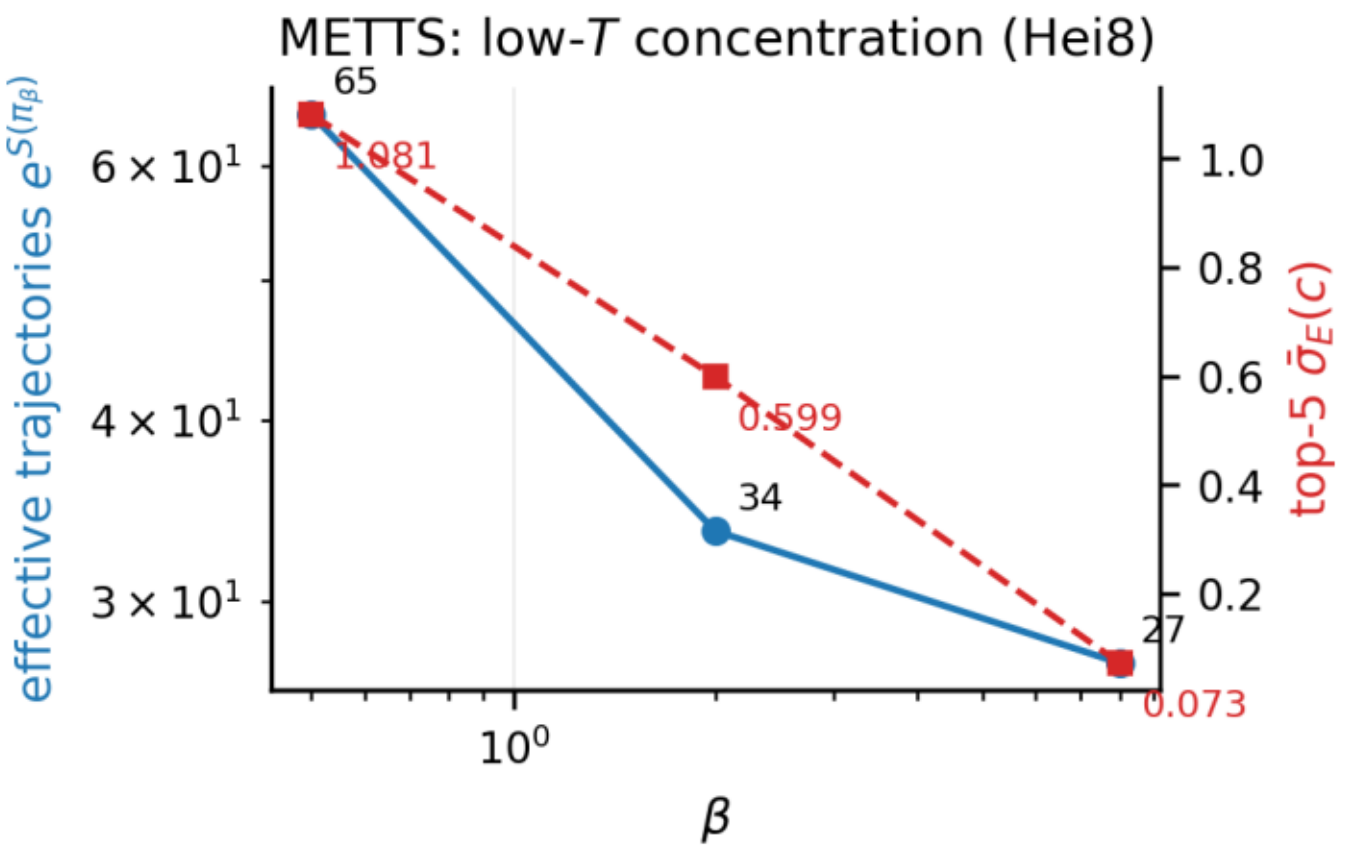

*Fig. 11. METTS low-temperature concentration. Heisenberg ring N = 8, full enumeration of the 70 collapse configurations: entropy participation ratio $e^{S(\pi_\beta)}$ (blue, left axis) and top-5 trajectory energy width $\bar\sigma_E$ (red, right axis) vs. $\beta$. Both fall with $\beta$, trajectories concentrate and sharpen ($\bar\sigma_E \approx 0.073 \approx 0.6/\beta$ at $\beta = 8$); while the METTS identity holds to $2 \times 10^{-14}$ at every $\beta$ (W2a). $e^{S(\pi_\beta)}$ is a participation ratio, not a sampling complexity; measured lag-1 autocorrelations are 0.96/0.51/0.01 (Sec. 5.4).*

## 5.5 The decision table

Table 3(a). Preparation-route decision table, capabilities (numeric anchors: Heisenberg ring N = 8, d = 70).

| route | prepares | readouts allowed | β-phase |
|---|---|---|---|
| (i) QSP direct | $\|\Phi_\beta\rangle = (e^{-\beta H/2} \otimes I)\ \|\Omega_c\rangle$ — exactly the Schmidt/TFD state | Δ ✓; F ✓; bin ✓ | $\sqrt{d/Z}$ can grow exponentially in system size at fixed $\beta$ (positive-entropy-density systems; joint $\beta \to 0$ limit relieves) |
| ($i'$) coherent thermal prep [7] | canonical purification to vector-norm error $\varepsilon$ (Thm. I.4) | Δ ✓ ($\varepsilon$ in budget); F ✓; bin ✓ | $\lambda_{gap}$-parameterized (rapidly thermalizing) |
| (ii) typicality (cTPQ) | $e^{-\beta H/2}\|\phi_{rand}\rangle$ (normalized filtered state) | bin ✓; F ✓ (single-copy correlators); Δ ✗ for the two constructions tested (Sec. 5.1; other constructions not excluded) | all temperatures, per-estimator ε-scaling (Lemma 8: $\varepsilon^{-2}-\varepsilon^{-3}$ depending on allocation); estimator-(a) carries a measured ensemble bias |
| (iii) METTS | trajectory states $e^{-\beta H/2}\|c\rangle$ | bin ✓ (adaptive); F ✓; Δ ✗ (trajectory states are not canonical purifications; not analyzed further) | low-T concentration (measured trend) |
| ground-state approx. (reference) | $\|\psi_0\rangle$ (zero-T circuit as-is) | zero-T machine | criterion of Cor. 6 met: no preparation needed |

Table 3(b). Preparation-route decision table; cost structure and measured parameters. All costs enter multiplicatively into the readout iterations (Theorem 10).

| route | cost structure ($U_H$ queries; multiplicative) | key parameters (measured) |
|---|---|---|

| route | cost structure ($U_H$ queries; multiplicative) | key parameters (measured) |
|---|---|---|
| (i) QSP direct | per window: $M_{\mathrm{AE}}(W)\cdot[2\sqrt{d/Z_{\mathrm{lb}}}\cdot d_{\mathrm{vt}}+4d_W]$ — with Z known (the anchors), $Z_{\mathrm{lb}}=Z$ and this is exact for this route (fixed amplified state, success probability exactly Z/d) — $d_{\mathrm{vt}}$ per Lemma 7 (opt $\tilde{O}(\sqrt{\alpha\beta\cdot\log(d/(Z\varepsilon_p))})$, cons column $O(\alpha\beta+½\ln(d/Z)+\ln(1/\varepsilon_p)))$ | d, Z: $\sqrt{d/Z}$ = 2.08/5.88/8.30 at $\beta$ = 0.5/2/8 |
| $(i')$ coherent thermal prep [7] | replaces $\sqrt{d/Z}\cdot d_{\mathrm{vt}}$ by $\tilde{O}(\beta^2\lVert H\rVert/\lambda_{gap}^{3/2}+\beta/(\varepsilon\cdot\lambda_{gap}^{3/2}))$ per prep call | their Table 1 (v2) |
| (ii) typicality (cTPQ) | per state per iterate: $2\cdot E[D^{-1/2}]\cdot d_{\mathrm{vt}}+4\cdot d_W^{\mathrm{bin}}$ (amplified; factor-level comparison $E[D^{-1/2}]$ measured 2.09/6.34/12.20 vs naive $\sqrt{d/Z}$ = 2.08/5.88/8.30 — Jensen); post-selection variant $E[D^{-1}]\cdot\alpha\beta$ (4.36/42.30/228.53 vs naive d/Z = 4.34 /34.57/68.90); iterates $M_{\mathrm{AE}}^{\mathrm{bin}}=O(\zeta_{\omega}/(\varepsilon_{\mathrm{rel}}J))$; states M = $O(\sigma_X^2/(\varepsilon_{\mathrm{rel}}J)^2)$, batched variant $M^{(b)}=O(\sigma_b^2/(\varepsilon_{\mathrm{rel}}J)^2)$ — multiplicative (Lemma 8) | $\sigma_X, M, b$: b/J = 1.96%/3.84% /1.65%; $\sigma_b/J$ = 0.189/0.180 /0.010; estimator (b) asymptotically unbiased, O(1/M) bias (T5b/T5i) |
| (iii) METTS | per trajectory per window: $M_{\mathrm{AE}}^{c}\cdot(2N_{\mathrm{prep,call}}^{c}+4\cdot d_W^{\mathrm{bin}})$, sampling complexity open | $e^{S(\pi)}$: 65.1/33.5/27.2 (participation ratio); $\rho_1$ = 0.96 /0.51/0.01 (T5e) |
| ground-state approx. (reference) | zero extra (Cor. 6) | G, $\lVert D\rVert^2$; $\beta^*(\varepsilon=0.05)$ = 20.46 on Hei8 |

Route F (any $\rho_\beta$ realization, per-window readout cost, no Δ-style preparation constraint): $\tilde{O}(\lVert c\rVert_1\cdot(\alpha/\gamma_W)\cdot\mathrm{polylog}(1/\varepsilon_F, 1/\varepsilon_{\mathrm{sim}}, 1/\varepsilon_{\mathrm{rel}}))U_H$ queries for the window quadrature (Sec. 5.1), same order as the duplex *readout* degree; preparation enters per the $\rho_\beta$ realization chosen (multiplicatively for coherent readout). Route F is the route of record when only mixed-state or sampling preparations are available and transition windows (not absolute windows) are the target.

Selection rules. (1) Check the ground-state criterion first (Cor. 6): if it passes, the zero-temperature circuit is already the thermal answer. (2) If not passed, the readout route decides the preparation class: the duplex Δ route requires a Schmidt/TFD-compatible preparation; route (i), or $(i')$ when its mixing-time assumptions hold; the Fourier-LCU route F delivers transition-window weights from any $\rho_\beta$ realization (mixed, typicality, METTS) at a per-window cost of the same order as the Δ readout degree (both $\tilde{O}(\alpha/\gamma_W)$-type up to polylogarithmic factors left unaccounted on the F side; the comparison of constants is system-dependent and not developed here); the bin route delivers absolute windows only. The duplex route's edge over F is the one-step sharp-indicator readout on a fixed preparable state, not exclusive access to transition windows. (3) Within the bin class at moderate temperatures, the direct-preparation routes are cheaper per our anchors (Sec. 6, Table 4): route (ii) pays $M\cdot M_{\mathrm{AE}}^{\mathrm{bin}}$ total AE iterates (each random state needs its own estimation), while route (i) re-prepares the *same* $|\Phi_\beta\rangle$ across its $M_{\mathrm{AE}}^{\mathrm{bin}}$ iterates ($\approx 15-17\times$ cheaper same-object at $\beta$ = 0.5/2); typicality becomes competitive at low temperature (M → 1 while route-(i) costs grow, Table 4) and remains the option when no purification-compatible route is available, with the estimator and its bias declared: estimator (a) requires $\varepsilon_{\mathrm{rel}}\gtrsim$ b/J, estimator (b) is asymptotically unbiased (O(1/M) bias) with batched readout. (4) Low temperature, large systems: all purification routes carry the $\sqrt{d/Z}$ factor per preparation call (exponentially growing in system size in positive-entropy-density systems at fixed $\beta$); METTS trajectory concentration and mixing-based preparations ($t_{mix}$-parameterized [7,11,13–15]) are the alternatives, with sampling complexities that we do not resolve here. (5) The full comparison is system- and window-dependent; the table carries the parameters, not a universal ranking.

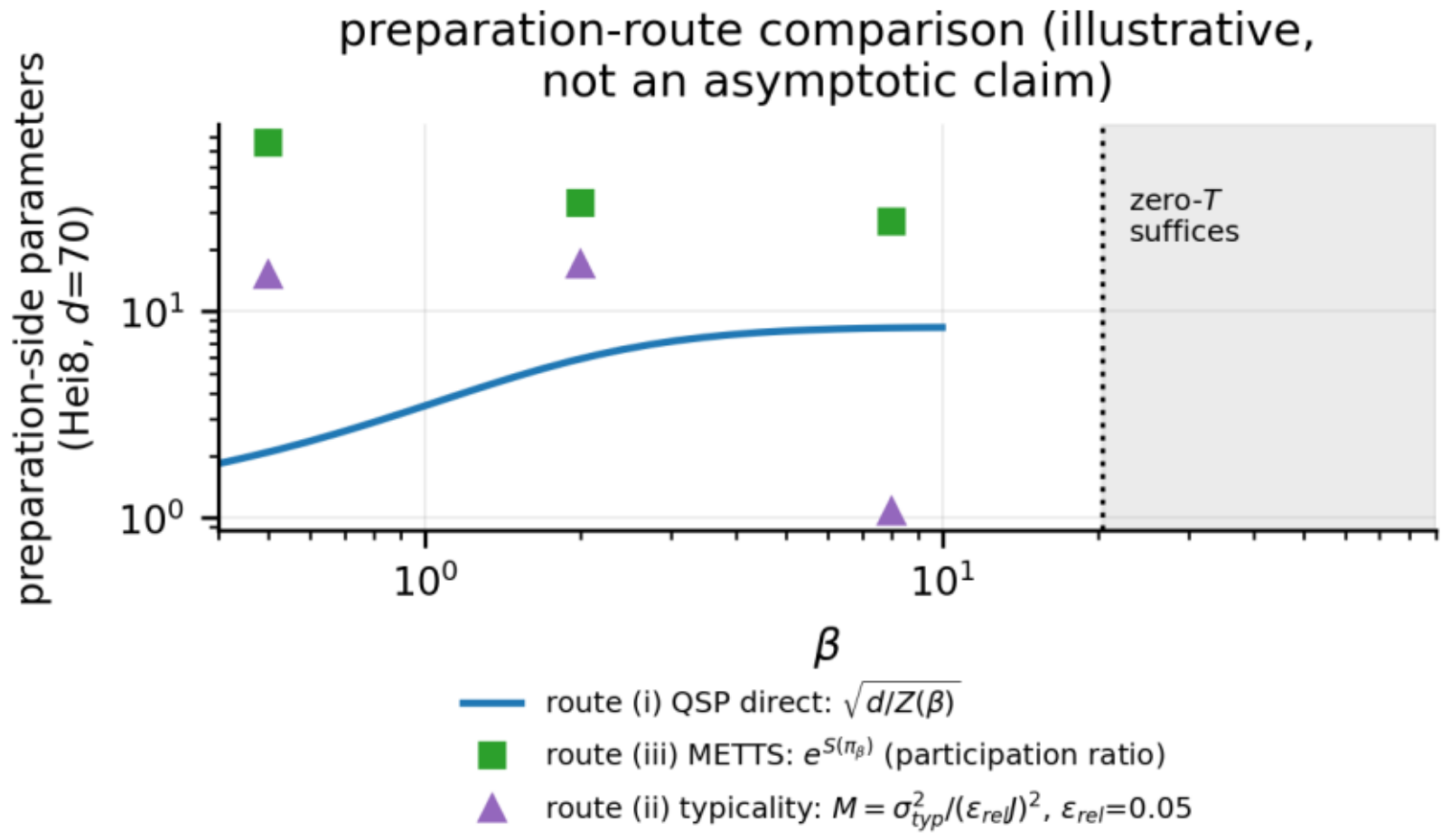

*Fig. 12. Preparation-route comparison. Heisenberg ring N = 8 (d = 70, G = 0.5227, $\|D\|^2$ = 16.0): preparation-side parameters vs. β (log–log). Blue curve: route (i) $\sqrt{d/Z(\beta)}$ (dense grid; 2.08/5.88/8.30 at β = 0.5/2/8; exact for route (i); the amplified state is fixed, with success probability exactly Z/d). Green squares: route (iii) entropy participation ratio $e^{S(\pi_\beta)}$ = 65.1/33.5/27.2. Violet triangles: route (ii) random-state count M = $\sigma^2_{typ}/(\varepsilon_{rel}\cdot J_\beta^{[0,3]})^2$ at $\varepsilon_{rel} = 0.05$ = 15.0/16.8/1.07; the target here is the bin-route absolute-window weight J (5.7025 at β = 2), not the Δ-route transition weight $S_\beta^{[1,38]} = 1.0809$. Route (ii)'s per-state preparation amplifier involves $E[D_\phi^{-1/2}](1.0025\times/1.0777\times/1.4699\times$ above $\sqrt{d/Z}$ at these β, T5i), which this axis does not display. Dotted vertical line: the ground-state criterion threshold $\beta = \ln(2(d-1)\|D\|^2/\varepsilon)/G = 20.46$ for $\varepsilon = 0.05$ (Cor. 6 bound with $\|D\|^2$); to its right (shaded) the zero-temperature circuit suffices. These curves compare parameters of different kinds (amplification factor, state count, participation ratio) on one small system; the figure is an illustration, not an asymptotic claim or a route ranking; full multiplicative cost comparisons are made in Sec. 6 (Table 4).*

---

# 6. The full query-complexity theorem

## 6.0 Accounting conventions

All cost statements follow eight accounting conventions, numbered 1–8 and stated in full in App. K.1 of the Supplemental Material: multiplicative preparation, readout assembly, window semantics (J vs. S), the window-edge condition, the Õ convention, window-polynomial realization, the input/output of Theorem 10, and the post-hoc ramp status. Individual statements cross-reference those item numbers.

Theorem 10 (thermal window weights; assembled, multiplicative accounting, a *conditional* query upper bound). Let $H_0$ have the $(\alpha, m, 0)$-block-encoding $U_H$ with $\alpha \geq ½$ (any block encoding can be rescaled to satisfy this) and D the $(\alpha_D, k, 0)$-block-encoding;

spectrum shifted $\lambda_0 = 0$, target: each of $N_w$ transition windows W to relative accuracy $\varepsilon_{\text{rel}}$, with $\mu_\beta(W) = S_\beta^{[a,b]}$, $Z = \sum e^{-\beta\lambda_j}$, $d = \dim$, $d_{\text{vt}} = \tilde{O}(\sqrt{\alpha\beta \cdot \log(d/(Z\varepsilon_p))})$ (optimal; Lemma 7, [39]; conservative column $O(\alpha\beta$ + ½log(d/Z) + log$(1/\varepsilon_p)))$, $d_W = O((2\alpha^{\text{eff}}/\gamma_w) \cdot \log^2(2/\varepsilon'))$, counted in UH-family queries, $2\alpha^{\text{eff}} = 2\alpha + |c|$, the routed SELECT of Sec. 4.1 making HΔ-query and UH-family-query counts identical.

Degree-bandwidth convention (stated once): $d_W$ is set by the *transition band* $\Delta = \gamma_w/O(\log(2/\varepsilon'))$; the band the ramp condition is stated on (App. B item 3): a degree-d polynomial resolves a normalized band $\approx \log(2/\varepsilon')/d$, so the second logarithm in $d_W$ is the cost of shrinking the band one logarithm below the window width. A degree-$(2\alpha^{\text{eff}}/\gamma_w) \cdot \log$ polynomial attains only sup ≈ 0.2 on the [1,38] anchor, so the $\log^2$ is required; the sup-error decay constant is degree-dependent ($\approx 3$ at small degree, drifting toward $\approx 1$ (the optimal-polynomial scale) at large degree), so constructions should carry a margin beyond the $\log^2$ formula value (T5k carries $\approx 25$; verification record in the supplemental material).

Then:

(a) Clean form (window-edge condition holds: no transition line of non-negligible thermal weight inside the QSP transition band of any W; checkable at design time when line positions are known, App. K.1 item 8); the total cost of route (i) + Δ (direct preparation + duplex readout) is at most

$N_{total} \leq \sum_W M_{\text{AE}}(W) \cdot ( 2 \cdot N_{\text{prep,call}} + 4 \cdot d_W$ + O(1) ), $N_{\text{prep,call}} = O(\sqrt{d/Z_{\text{lb}}}) \cdot d_{\text{vt}}(Z_{\text{lb}} = a$ known lower bound on Z; $Z_{\text{lb}} = Z$ when Z is known, $Z_{\text{lb}} \geq 1$ always, Lemma 7),

$M_{\text{AE}}(W)$ = O( max( $\sqrt{(1-p_W)/p_W}/\varepsilon_{\text{rel}}, 1/\sqrt{\varepsilon_{\text{rel}}}$ ) · $\log(1/p_{\text{fail}})$ ) (tight, signal-adaptive; $p_W = \mu_\beta(W)/\alpha_D^2$; the second term is the $p \to 1$ guard from the $\pi^2/M^2$ addend of the AE bound, negligible at our anchors, where the first term $\sqrt{(1-p)/p}/\varepsilon_{\text{rel}} \geq 74$ on the Δ rows (≥ 53 including the bin anchor)) ≤ O( $\zeta_\beta \cdot \log(1/p_{\text{fail}})/(\varepsilon_{\text{rel}} \cdot \mu_\beta(W))$ ) (safe upper bound; $\zeta_\beta = 2\alpha\alpha_D^2$), with the preparation unitary entering every coherent iterate of amplitude estimation (two calls per Grover iterate: A and $A$ †) and the conjugate-pair window realization applying two window sequences per chain $4 \cdot d_W$ per iterate, the exact-reflection readout circuit of App. C. In Õ form (the sum display, every route comparison and table of this paper refers to it):

$N_{total} \leq \sum_W$ O( $\log(1/p_{\text{fail}})$ ) · [ $\max(\sqrt{(1-p_W)/p_W}/\varepsilon_{\text{rel}}, 1/\sqrt{\varepsilon_{\text{rel}}}) \cdot \sqrt{d/Z_{\text{lb}}} \cdot \sqrt{\alpha\beta \cdot \log(d/(Z_{\text{lb}}\varepsilon_p))} + \max(\sqrt{(1-p_W)/p_W}/\varepsilon_{\text{rel}}, 1/\sqrt{\varepsilon_{\text{rel}}}) \cdot (2\alpha^{\text{eff}}/\gamma_w) \cdot \log^2(2/\varepsilon')$ ] (Eq. 10-sum; the max carries the $p \to 1$ guard, in the square-root-dominated regime ($\sqrt{(1-p)/p}/\varepsilon_{\text{rel}} \geq 1/\sqrt{\varepsilon_{\text{rel}}}$, true at all our anchors), it reduces to the previous display)

The exact per-iterate quantity is the sum $2 \cdot N_{\text{prep,call}} + 4 \cdot d_W$; the former *product* display $\tilde{O}(\sqrt{d/Z} \cdot d_{\text{vt}} \cdot (2\alpha^{\text{eff}}/\gamma_w) \cdot \zeta_\beta/(\varepsilon_{\text{rel}} \cdot \mu))$ is its product-form upper bound (a sum bounded by a product; retained for backward comparability and always strictly ≥ the sum form). The tight and safe $M_{\text{AE}}$ columns differ substantially at small signals: on the Hei8 anchors the ratio $\zeta_\beta/\mu \div \sqrt{(1-p)/p}$ is 51/45/179 at $\beta$ = 0.5/2/8 (Table 4), the safe column systematically overcharges small-signal windows, and route comparisons in this paper quote both.

(b) General form (no line-free assumption): the accuracy guarantee of (a) acquires the explicit additive error term

$|\hat{\mu} - \mu_\beta(W)|/\mu_\beta(W) \le \varepsilon_{\mathrm{rel}} + C_{\mathrm{ramp}} \cdot w_{\mathrm{ramp}}(W)/\mu_\beta(W), C_{\mathrm{ramp}} \le 1, w_{\mathrm{ramp}}$ = thermal line weight inside the transition band,

at the same cost, with $w_{\mathrm{ramp}}$ measurable post hoc by the edge-adjacent windows of the same plan (App. K.1 item 8, T5h quantifies it on the [1,38] anchor: $w_{\mathrm{ramp}}/\mu_W(\delta = 0.05)$ reaches 21% at $\beta = 2$ and 31% at $\beta = 0.5$ (73.8% at $\beta = 0.5, \delta$ = 0.25); the anchor budget there is ramp-limited; T5k supplies a line-free anchor where (a) closes).

(Unknown signal $\mu_\beta(W)$ (protocol, Õ-level sketch): run AE at iterate counts M = $2^3, 2^4, 2^5, \ldots$ (geometric doubling). At level M the AE output $\hat{p}_M$ carries the BHMT confidence interval $\hat{p}_M \pm (2\pi\sqrt{\hat{p}_M(1-\hat{p}_M)}/M + \pi^2/M^2)$; the schedule stops at the first M whose interval has relative half-width ≤ $\varepsilon_{\mathrm{rel}}$. The total iterate count is the geometric sum ≤ $2 \cdot M_{final}$, with per-level failure probability $p_{\mathrm{fail}}/2^k$ (union bound preserving the overall $1 - p_{\mathrm{fail}}$), and the returned estimate is the standard AE readout at $M_{final}$ (no post-processing beyond the interval check); $M_{final}$ equals the tight $M_{\mathrm{AE}}$ of (a) whenever the interval at that M is honest. The protocol needs no prior knowledge of $\mu$ beyond $p_W > 0$ (an absolute-precision fallback at $\tilde{O}(\alpha_D^2 \cdot (2\alpha^{\mathrm{eff}}/\gamma_w) \cdot \log^2(1/\varepsilon')/\varepsilon_{abs})$ queries covers the near-zero-signal case at absolute accuracy $\varepsilon_{abs}$).

*Route (ii) + bin (typicality):* Cor. 10.3 (per-state preparation multiplicative exactly as above; the ε-scalings of Lemma 8 are allocation-specific, Sec. 5.3).

*Route (iii) (METTS):* not part of this theorem. The METTS estimator's sampling theory (trajectory count, autocorrelation, burn-in) is open; the route is presented as an open preparation/readout direction with an exact identity and a cost *structure* in Sec. 5.4 (Lemma 9, App. A.10).

*Reference (zero-temperature, base paper Thm. 2 in its SNR-parametrized form, App. C, assembled with the same exact-reflection convention):* under the same multiplicative convention the zero-temperature readout of window W costs $M_{\mathrm{AE}}^0(W) \cdot (2 \cdot N_{\mathrm{prep}}^0 + 4 \cdot d_W^0), N_{\mathrm{prep}}^0 = O((\alpha/(\sqrt{p_0} \cdot G)) \cdot \log(1/\varepsilon))$ [assumption A1], $d_W^0 = O(((\alpha + |c|)/\gamma_w) \cdot \log^2(2/\varepsilon'))$ under the same off-center shift convention and degree-bandwidth convention as restated for this paper's machine (the base paper's original degree statement carries only a single logarithm under its own band convention, the paired accounting of Cor. 10.1; a genuinely centered zero-temperature window has c = 0 and $d_W^0 = O((\alpha/\gamma_w) \cdot \log^2(2/\varepsilon'))$); the additive "prepare once" presentation of [1] corresponds to a persistent-state reading that we do not invoke for either algorithm.

*Proof (assembly).* Object layer (Theorem 3 with the $(2\alpha, m+1, 0)$-encoding, one $U_H$-family query per $H_\Delta$ query via the routed SELECT); readout machine (Theorem C.1, whose iterate count gives the displayed $M_{\mathrm{AE}}$); preparation (Lemma 7); multiplicative accounting (each coherent iterate re-prepares and applies the conjugate pair); error budget (App. B). Step-by-step assembly, the per-constant bookkeeping, and the degree-sparing variant: App. J (with App. B/C).

Cor. 10.1 (readout-stage ratio vs. zero temperature; conventions must be paired (including the shift). At equal μ-normalization and *consistent off-center handling*) both machines realize a non-center window [a,b] by the same shifted-LCU construction with c = (a+b)/2 (the zero-temperature machine's window [a,b] ⊂ $[0,\alpha]$ is equally off-center and equally subject to the QSP parity constraint, so under consistent accounting it carries the same shift: $d_W$ vs. $d_W^0 = O(((\alpha + |c|)/\gamma_w) \cdot \log^2(2/\varepsilon')))$, the filter+AE stage of the Δ route exceeds the zero-temperature one by the filter-degree ratio

$(2\alpha + |c|)/(\alpha + |c|) = 1.79$ at the [1,38] anchor $(2\alpha^{\mathrm{eff}} = 12.80, c = 1.5)$ under tight normalization (subnormalization ratio 1 + $O(\varepsilon')$).

The often-quoted $2\alpha^{\mathrm{eff}}/\alpha \approx 2.27$ arises only by charging the thermal machine the shift while leaving the zero-temperature window centered (a mixed convention; legitimate only when the zero-temperature target is a genuinely zero-centered window); the naive two-instance SELECT gives 4.53; the safe-normalization counterparts (composite-subnormalization ratios $\zeta_\beta/\zeta_\omega$ with $\zeta_\beta = 2\alpha\alpha_D^2$) are 1.58 (zero-T machine shifted, $\zeta_\omega^{\mathrm{shift}} = (\alpha + |c|)\alpha_D^2$, thermal machine at its base normalization) and 1.39 (zero-T centered, $\zeta_\omega = (\alpha + 1 + \bar{\omega})\alpha_D^2$, the archived t5d convention); under safe normalization the thermal numerator stays $2\alpha$, so the fully shift-paired safe ratio coincides with the tight one only in normalization-1 accounting; the definitions are archived with the values. All five conventions, with their definitions, are archived in the supplemental verification records (`readout_ratios` section; the centered-convention quartet 2.265 /2.773/4.0/5.546 is listed for comparison). Any accounting that gives both machines the same exact-reflection assembly (conjugate pair in A and $A$ †) and the same shift convention is a paired convention; we quote the paired-and-shifted value 1.79 as primary. Larger factors (e.g. 4.53 under the naive two-instance SELECT) arise under constructions eliminated by the routed SELECT. Assumption A1 $(p_0, G)$, does not appear anywhere in the thermal readout: the preparation parameters $\sqrt{d/Z_{\mathrm{lb}}} \cdot d_{\mathrm{vt}}$ or $e^{S(\pi_\beta)}$ are system/temperature parameters, not spectral-gap parameters; though at low temperature $\sqrt{d/Z}$ grows and can dominate whatever A1 would have cost (Sec. 8).

Cor. 10.2 (conditional self-checks). $N_w$ windows each carry their own full per-window cost, preparation included (Theorem 10; the preparation is re-paid per window, not amortized). FDT-L mirror consistency $(S^L(\omega) vs. S^L(-\omega))$ and weight-level detailed balance cost zero additional queries if the mirrored negative-frequency windows are already in the measurement plan; a mirrored window outside the plan costs its own readout $(M_{\mathrm{AE}}(-W) \cdot (2N_{\mathrm{prep,call}} + 4d_W))$; "zero additional circuit design" (the same circuit with a shifted window) is not by itself "zero additional queries".

Cor. 10.3 (typicality route). Route (ii)+bin totals $M \cdot M_{\mathrm{AE}}^{\mathrm{bin}} \cdot (2 \cdot N_{\mathrm{prep}}^{\mathrm{state}} + 4 \cdot d_W^{\mathrm{bin}})$ with M = $O(\sigma_X^2/(\varepsilon_{\mathrm{rel}}J)^2)$ and $M_{\mathrm{AE}}^{\mathrm{bin}} = O(\zeta_\omega \cdot \log(1/p_{\mathrm{fail}})/(\varepsilon_{\mathrm{rel}}J))$; the ε-scalings of Lemma 8, which are allocation-specific upper bounds (per-state AE error treated as systematic at $\Delta = \varepsilon_{\mathrm{rel}} \cdot J$; an independent-error allocation reaches $\varepsilon^{-2} - \varepsilon^{-2.5}$ (Sec. 5.3/Lemma 8), preparation multiplicative exactly as in route (i)) where J is the bin-route absolute-window weight ($J_\beta^{[0,3]}$ in our anchors) and the amplified per-state preparation is $N_{\mathrm{prep}}^{\mathrm{state}} = O(E[D_\phi^{-1/2} \cdot d_{\mathrm{vt}}(\phi)])$, the expectation of the product (Lemma 8's primary form); the ensemble anchor form $E[D_\phi^{-1/2}] \cdot d_{\mathrm{vt}}$ (= 21.8/126.6/661.7 at $\beta$ = 0.5/2/8, conservative $d_{\mathrm{vt}}$ column) approximates it with measured coupling excess 0.03%/0.51%/1.07% at $\beta$ =

0.5/2/8 (T5i), and the naive $\sqrt{d/Z} \cdot d_{\mathrm{vt}}$ is a Jensen lower bound (measured gap up to 1.4699 × at $\beta = 8$, Lemma 8; both gaps measured on d = 70, Sec. 5.3); the post-selection variant $O(E[D_{\phi}^{-1}] \cdot \alpha\beta)$ (measured gap up to 3.3167 ×) is measurement-based and fits only shot-style readout. The per-state ratio estimator carries the measured ensemble bias b/J = 1.65–3.84% (estimator-(a)-specific); the batched estimator (b) is asymptotically unbiased (O(1/M) bias, under the moment conditions of App. A.9) with its own variance parameter, $M^{(b)} = O(\sigma_b^2/(\varepsilon_{\mathrm{rel}} J)^2)$,$\sigma_b/J$ measured 0.189/0.180/0.010, the variance law consolidated at 40 batches for all three $\beta$ (T5b-ext; ratios 1.13–1.16); and its quantitative circuit accounting beyond the M-formula is left at the protocol-sketch level (Sec. 5.3).

Cor. 10.4 (ground-state criterion). When $\beta G \geq \ln(2(d-1)\|D\|^2/\varepsilon)$ (Cor. 6; sufficient condition), the zero-temperature circuit of the base paper outputs the thermal answer to accuracy $\varepsilon$, and METTS is a candidate interpolation across the criterion (Lemma 9's measured concentration trend; no sampling theorem is claimed).

Table 4. Multiplicative-accounting anchors, dual-convention (sq ≡ $\sqrt{d/Z}$; Hei8, $\varepsilon_{\mathrm{rel}} = 0.05$; illustrative units $\alpha = 5.651$, $\alpha_D = 4$, $\gamma_w = 1$, $\varepsilon_p = \varepsilon' = 10^{-3}$; the Δ-route and bin/typicality columns target different objects ($S_\beta^{[1,2]}$ vs. $J_\beta^{[0,3]}$) so cross-object columns are starred; $M_{\mathrm{AE}}$ in two conventions (tight $\sqrt{(1-p)/p}$, safe $\zeta/(\varepsilon_{\mathrm{rel}}\mu)$) and $d_{\mathrm{vt}}$ in two columns (cons/opt); the tabulated $M_{\mathrm{AE}}$ omits $\log(1/p_{\mathrm{fail}})$ and charges AE at the App.-B $2\varepsilon_{\mathrm{rel}}/3$ allocation; the anchor windows are ramp-limited at $\beta \leq 2$ (T5h), arithmetic illustrations under declared units, not resource estimates; the ramp-valid anchor is the T5k row. Full note: App. L.

| $\beta$ | sq | $d_{\mathrm{vt}}$ cons | $d_{\mathrm{vt}}$ opt | $M_{\mathrm{AE}}\Delta$ tight / safe | $M_{\mathrm{AE}}^{\mathrm{bin}}$ tight / safe | $N^{(i,\Delta)}$ t+o | $N^{(i,\Delta)}$ s+c | $N^{(ii)}$ t+o | ii / Δ t+o | ii / bin t+o | ii / bin s+c |
|---|---|---|---|---|---|---|---|---|---|---|---|
| 0.5 | 2.08 | 10.5 | 4.6 | 128 / 6467 | 80 / 1983 | $3.8 \times 10^5$ | $1.9 \times 10^7$ | $2.7 \times 10^6$ | 7.1 | 15.0 | 15.1 |
| 2 | 5.88 | 20.0 | 9.9 | 111 / 5019 | 40 / 686 | $3.4 \times 10^5$ | $1.6 \times 10^7$ | $1.6 \times 10^6$ | 4.7 | 16.9 | 17.1 |
| 8 | 8.30 | 54.2 | 20.2 | 473 / 84742 | 34 / 564 | $1.6 \times 10^6$ | $3.3 \times 10^8$ | $1.0 \times 10^5$ | 0.064 | 1.14 | 1.24 |

($S_\beta^{[1,38]}$ = 0.839/1.081/0.064; $J_\beta^{[0,3]}$ = 1.973/5.702/6.936; $d_W^{\Delta} = 740$, $d_W^{\mathrm{bin}} = 558$ under the degree-bandwidth convention ($2\alpha^{\mathrm{eff}} = 12.80$ for the [1,38] window; shift bound $\alpha + E_{\mathrm{cut}}$, $E_{\mathrm{cut}} = 4$; log-degree values such as 98/74 underresolve the declared transition band, sup ≈ 0.2 (verification record in the supplemental material)), both × 4 per iterate, tight/safe $M_{\mathrm{AE}}$ relaxation ratios: 51 ×/45 ×/179 × at $\beta$ = 0.5/2/8, the safe convention systematically overcharges small-signal windows, most strongly at low temperature; the $d_{\mathrm{vt}}$ opt column is ≈ 2 − 2.7 × below cons; *the $N^{(ii)}/N^{(i\Delta)}$ column compares different window semantics (J vs. S). The earlier single-convention (safe+cons) table is recoverable from the safe+cons column; all rows are arithmetic illustrations, not resource estimates.)

**Ramp-valid anchor (T5k), dual-convention.** The window $W = [2.431, 2.467]$ at $\beta = 2$ (isolated line $\omega = 2.4495$, weight 0.326) satisfies the window-edge condition of App. B item 3 at the budget's own bandwidth, counted two-sidedly: the in-band weight share is $2.6 \times 10^{-33}$, and the nearest weighted line sits 1.43× the band away. With the App.-B budget allocations ($\varepsilon_p = 8.55 \times 10^{-5}$, $\varepsilon' = 2.25 \times 10^{-4}$), $2\alpha^{\mathrm{eff}} = 13.75$, $d_W \approx 4 \times 10^4$ (construction margin), $S = 0.3284$, $d_{\mathrm{vt}}$ cons/opt = 22.44/11.22, MAE tight/safe = 207 /16522 (relaxation 80×), the four-term budget closes at $\varepsilon_{\mathrm{rel}}$: **N**$_{\mathbf{total}}$ **=** $3.3 \times 10^7$ **(tight+ opt) /** $2.6 \times 10^9$ **(safe+cons)**. Two rejected candidate windows and the full arithmetic: App. L.

Reading (illustrative arithmetic under the stated conventions, one $d = 70$ system, not resource estimates, not a route ranking): (1) the multiplicative-vs-additive correction is mild at high temperature and moderate at low (≈ 1.01 ×/1.08 ×/1.30 × at $\beta = 0.5/2/8$); (2) same-object, the direct-preparation route is cheaper by 15–17× at $\beta = 0.5/2$, and at $\beta = 8$ the same-object ordering narrows to 1.1–1.2×, while the cross-object column (J vs. S) moves from 5–7× to 0.064; **the crossover direction and size are convention- and object-dependent**, so no ranking is drawn; (3) typicality's precision is capped by the measured ensemble bias ($\varepsilon_{\mathrm{rel}} \gtrsim$ 1.65–3.84% for estimator (a) on this system); (4) the tight-vs-safe MAE gap (45–179×) is itself the largest single uncertainty in any route arithmetic.

Assembly status. Theorem 10 is an *assembled* theorem: the components (object identity Thm. 3, readout machine Thm. C.1, preparation Lemma 7 and the exact identities of Lemma 8) are individually verified, and the assembly's per-constant consistency is asserted at Õ level (App. B expands the logarithmic factors and assembles the error budget). Literature preparation routes' exponents enter as black-box parameters ($t_{mix}$, Z-type bounds), with the specific constants quoted from the sources listed in Sec. 2.

---

## 7. Numerical validation

All experiments use the lattice exact-diagonalization module of the Supplemental Material (deterministic; seed 20261023 wherever random states are drawn). The systems are the Heisenberg ring at N = 8, 12, 16 (d = 70, 924, 12870; the last truncated to k = 256), the Hubbard 2×2 plaquette (dim 36), and the Hubbard dimer (dim 6); Hamiltonians, dipoles, shifted spectra ($\lambda_0 = 0$), atomic units, and all other conventions are listed in App. L and Table D.5 of the Supplemental Material.

Numerical verification is organized in two tiers, and the text refers to individual checks by label. Tier A (algebraic PASS) covers the V/P/W/N/T5a–T5g checks: different formulas and different matrix constructions are evaluated on the same exact-diagonalization eigenpairs (they share one ED spectrum, so "independent" means independent formula paths and contractions, not independent spectra). Tier B (matrix-host PASS) covers tp6 C2–C4, which validate the readout assembly as explicit unitaries on the symmetric-dilation host, under the block-encoding promise: $A_\beta$ enters at the exact-unitary level, and Lemma 7's internal QSP structure is not re-verified there. Validation is matrix-level under the block-encoding promise; the untested items are listed in Sec. 8.2. Every "circuit-level validation" statement in this paper means Tier B, and "one H-family query per $H_\Delta$ query" is a query-model statement, not a gate count. The complete ledger, with every check, its systems, criteria, and PASS statement, including the registered exceptions and the single-run archive values, is Table E.1 of the Supplemental Material.

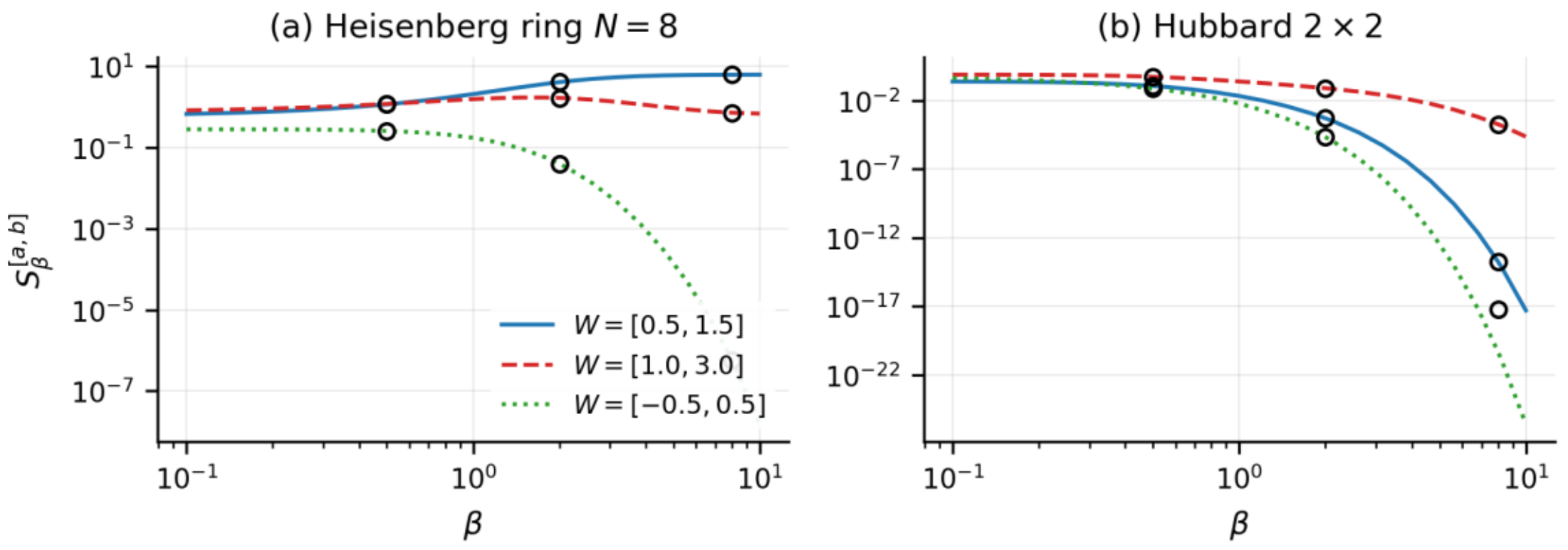

*Fig. 13. Window weights vs. temperature. Transition-window weights $S_\beta^{[a,b]}$ for windows [0.5,1.5], [1,8], $[-0.5,0.5]$ on (a) the Heisenberg ring N = 8 and (b) the Hubbard $2 \times 2$ plaquette, $\beta \in$ [0.1,10] (log–log). Hollow markers: at every $\beta \in$ {0.5, 2, 8} on all six curves, independent dense-$H_\Delta$ constructions (explicit Kronecker difference operator, 4900/1296-dimensional eigendecomposition, no use of the $\lambda_n - \lambda_m$ structure) reproduce the scan values, 18 anchors in total, relative agreement ≤ $7.02 \times 10^{-10}$ on signaled cases (absolute ≤ $6.2 \times 10^{-18}$ on two near-zero cases). The zero-temperature limit (right edge) and the high-temperature plateau (left edge) bracket order-of-magnitude redistributions, e.g. Hei8 [0.5,1.5] grows from 1.152 ($\beta = 0.5$) to 6.190 ($\beta = 8$) as the ground-state channel takes over (the $\beta = 8$ value coincides with the absolute-window value 6.127 up to the $e^{-\beta G}$-scale residual excited-state weight, as it must at low temperature).*

*Highlights (identities at two independent formula paths, the FDT-L check, the sampling/truncation behaviour) are collected in App. J of the Supplemental Material; the archived values are those of Table E.1.*

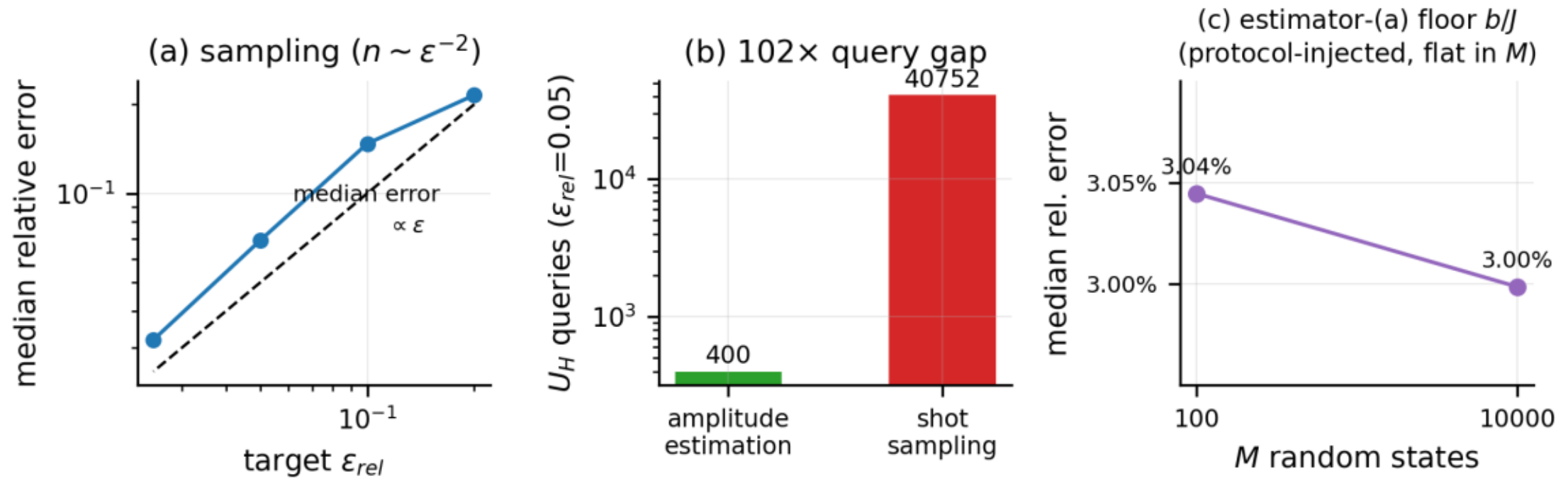

*Fig. 14. Sampling-protocol scaling. Hubbard* $2 \times 2, \beta = 2$, *window [1,2]. (a) Shot-noise simulation (60 experiments per point): median relative error vs. target* $\varepsilon_{rel}$ *tracks the diagonal (n ~* $\varepsilon^{-2}$*). (b) Queries at* $\varepsilon_{rel} = 0.05$*: analytic amplitude-estimation query count 400 vs. binomial shot-noise simulation 40752, not a common-circuit experiment (the two sides use different accounting bases: AE queries are the Õ formula's count for the nominal signal, the shot side is a Monte Carlo of the sampling protocol; the same-amplitude same-accounting version of this comparison is tp6-C4, Sec. 7.1, which evaluates the gap on one assembled circuit (analytic sampling models of its p on both sides)). (c) The estimator-(a) ensemble bias injected into the sampling protocol: median error stays ≈* 3 *(3.04% at M = 100; 3.00% at M =* $10^4$*); flat in M, the signature of the per-state-ratio estimator's systematic bias b (estimator-specific: the ratio-of-means estimator of Lemma 8 removes it).*

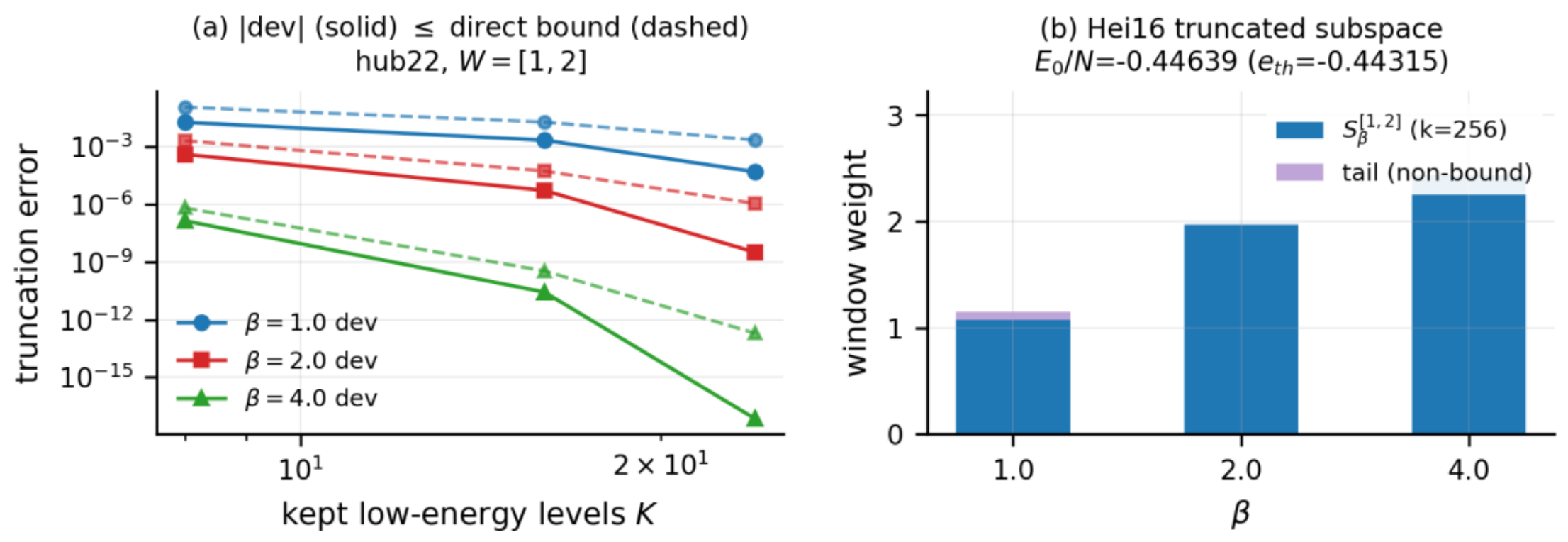

*Fig. 15. Low-energy truncation. (a) Hubbard* $2 \times 2$*: truncation deviation (solid) vs. the direct tail bound (dashed) for β ∈{1,2,4}, kept levels K ∈{8,16,24}; the bound holds in all nine combinations. (b) Heisenberg ring N = 16 (dim 12870, k = 256 kept): window weight* $S_\beta^{[1,38]}$ *and the geometric tail estimate (non-bound; estimate only) for β ∈{1,2,4};* $E_0$*/site =* $-0.44639$ *vs. thermodynamic* $-0.44315$.

## 7.1 Circuit-level spot verification (tp6; matrix-level simulation)

**Level statement.** The check is *matrix-level*: every unitary (the pyqsp window sequence, the dipole dilation, the exact reflections, the Grover iterate) is an explicit matrix; it is not gate-level synthesis, and the preparation enters at the exact-unitary level. The $H_\Delta$ block encoding used inside C2–C4 is the symmetric dilation of $H_\Delta/(2\alpha)$, while the routed SELECT is validated separately in C1 to the same top-left block. All four checks pass (C1 top-left block to $5.6 \times 10^{-17}$; C2 coin identity to $1.4 \times 10^{-13}$; C3 exact-reflection AE cluster to $2.1 \times 10^{-15}$; C4 same-amplitude AE-versus-shot 4.3× at the tested amplitude). The 4×4 Hubbard validation was not performed, and we say so; the N = 16 Heisenberg truncated run substitutes methodologically. *Per-check detail, the realization notes, and the noted divergence: App. I.*

# 8. Discussion

## 8.1 Structural lessons

1. Absolute vs. transition energies. The zero-temperature framework's innocuous-looking shift $\lambda_0 = 0$ is precisely what fuses "circuit-measurable spectral functions" with "response-theoretic transition frequencies". At finite temperature the fusion breaks, and every algorithmic object must declare which axis it lives on (Eq. 2 vs. Eq. 1, and, within the bin class, J vs. S; Sec. 5.3). We expect this axis discipline to matter in every thermal extension of filtering-based quantum algorithms.
2. Purification conventions are load-bearing. The duplex object exists only as an expectation value on a *specific* purification (Schmidt/TFD canonical); cyclic traces compute a different (density-of-states-type) object, and non-conjugate ancilla rotations change the answer. That two concrete preparation routes (direct QSP; the coherent algorithm of [7]) produce exactly the required state is what makes the duplex route coherent end-to-end.
3. Preparation and readout must be co-designed; and preparation is not amortizable. The compatibility matrix (Sec. 5.1) is not an afterthought: choosing typicality for preparation forfeits the duplex readout for the two constructions tested (other typicality constructions are not excluded, Sec. 5.1), and choosing duplex readout restricts the preparation class. Under multiplicative accounting (Theorem 10), the preparation choice also enters every readout iterate, so "which preparation" is a per-iteration multiplier, not a one-time overhead; this changes route comparisons qualitatively (Table 4).

## 8.2 Scope of the results

The results above hold under the following scope.

(i) Theorem 10 holds as a query upper bound when the window edges are line-free; when they are not, the explicit $C_{\mathrm{ramp}} \cdot w_{\mathrm{ramp}}$ term (form b) is carried. Its iterate count appears in tight and safe conventions whose gap (45–179 × on our anchors) is the largest single uncertainty in the accounting, and the unknown-signal protocol (c) is a sketch. The components, Theorem 3, Theorem C.1 and Lemmas 7–9, are individually verified; App. B assembles the error budget at Õ level. On the circuit side, the routed-host AE stack, the DFT/QFT readout circuit, gate synthesis and fault-tolerant resource estimates are not tested; C1 validates the routed SELECT separately, to the same top-left block. A nondestructive multi-window readout would restore additive amortization. The safe normalization $\zeta_\beta = 2\alpha\alpha_D^2$ carries slack $O(\alpha)$ (Cor. 10.1), and literature preparation exponents enter as black-box parameters.

(ii) The validation covers dim924 exactly (Hei12), dim12870 truncated (Hei16), the 2 × 2 Hubbard plaquette and the dimer. The 4 × 4 Hubbard K = 0 sector (dimension 10,353,252) exceeds the dense-ED budget and is left to tensor-network/DMRG. The complex-conjugation conventions of Sec. 3.1 (and the $U_H{}^*$ branch of the routed SELECT) are pinned theoretically but not exercised numerically; a flux-threaded Hubbard ring is the natural first test.

(iii) Typicality's $\varepsilon$-scalings are allocation-specific ($\varepsilon^{-2}$–$\varepsilon^{-3}$; Lemma 8), and per-state preparation is governed by the negative moments of the normalization variable (measured Jensen gaps up to $1.47\times$ and $3.32\times$ at $\beta = 8$; T5i). The per-state ratio estimator carries a measured ensemble bias (1.65–3.84%; two-component fits ≤ 3.4%) with a Jensen-type mechanism; the ratio-of-means estimator is asymptotically unbiased and measured plateau-free (T5b, T5b-ext). $\sigma_X$ enters as a measured input, with no analytic law for $\sigma_X(\beta,\text{window})$. METTS concentrates at low temperature, $e^{S(\pi_\beta)}$ is a participation ratio rather than a sampling complexity, and the sampling overhead is open (lag-1 autocorrelations 0.96/0.51/0.01, T5e).

(iv) Window weights are estimated to relative accuracy $\varepsilon_{\text{rel}}$ (Theorem 10). The pointwise response follows on the same machinery: the damped-response objects are spectral functions of $H_\Delta$, and the Lorentz kernel $L_\gamma(H_\Delta - \omega)$ (no edge-line condition) yields $\text{Im}\chi_\beta$ directly by QSP on the same register, with the real-part kernel readable through a linear combination. This paper stays with the indicator-window primitive; the per-point cost analysis of the Lorentz/real-part variant and the adaptive window-selection protocol are left to follow-up.

(v) Higher-order response lives on the same $H_\Delta$ register: in the Liouville formulation (Sec. 4.4) the n-th-order denominators are functions of $H_\Delta$, and repeated application with dipole-side insertions $\mathcal{D}$^× is the doubled-register image of the standard double-sided Feynman expansion [2]. The construction and its resource accounting are left to future work.

## 9. Conclusion

The excite–filter–readout framework extends to finite temperature, and the extension is complete at the level of architecture, readout, and resource accounting. What the circuit estimates is a transition-window weight, the thermally weighted spectral content that supports the response; the architecture carries over, the resource logic does not. Theorem 3 is the Liouville-space/TFD superoperator projection made explicit in the block-encoding/QSP setting [2–6], and what we build around it is a routed-SELECT block encoding of the difference operator (one H-family query per filter query), a conjugate-pair readout of the real window polynomial, a four-route preparation–readout compatibility analysis (duplex Δ, bin, Fourier-LCU, METTS adaptive-bin), and a multiplicative resource accounting.

Theorem 10 prices a window at $M_{\text{AE}} \cdot (2\sqrt{d/Z_{\text{lb}}}\, d_{\text{vt}} + 4d_W)$ queries; at the anchor and under paired shift conventions the readout stage costs a factor $(2\alpha + |c|)/(\alpha + |c|) \approx 1.8$ more than at zero temperature (Cor. 10.1). Each coherent readout iteration re-prepares the state and applies the conjugate-pair window realization, at $O(\sqrt{d/Z_{\text{lb}}}) \cdot \tilde{O}(\sqrt{\alpha\beta \cdot \log(d/(Z_{\text{lb}}\varepsilon_p))})$ queries per preparation call on the direct route ($Z_{\text{lb}} = Z$ when Z is known). The temperature-dependent part is the preparation factor $\sqrt{d/Z}$: it grows exponentially at fixed temperature in positive-entropy-density systems, at rate $s(\beta) - (-\beta f(\beta))$ per site, with measured 2.08 → 25.7 from N = 8 to 14 at $\beta$ = 2, and is relieved only in the joint $\beta \to 0$ limit. Coherent thermal-state algorithms replace it by mixing-time parameters. On the readout side, the tight signal-adaptive iterate count $\sqrt{(1-p)/p}/\varepsilon_{\mathrm{rel}}$ carries the small-signal regime (the safe $\zeta_\beta$ form overcharges it by $45-179\times$ on our anchors), and the ramp term stays an explicit additive error unless the window edges are line-free.

Typicality behaves differently again. Its ε-scalings are allocation-specific ($\varepsilon^{-2}-\varepsilon^{-3}$ depending on the error allocation; Lemma 8), and per-state preparation is governed by the negative moments of the normalization variable (measured Jensen gaps up to $1.47\times$ and $3.32\times$ at $\beta=8$ on d = 70, small-system numbers whose $CV^2$ participation bound decays with system size). Two estimators are available. METTS concentrates at low temperature. The most structural change is in assumption A1: the ground-state overlap $p_0$ and gap G no longer enter the readout, only explicit preparation parameters do, subject to the sufficient criterion $\beta G \geq \ln(2(d-1)\|D\|^2/\varepsilon)$. Circuit-level statements are matrix-level validations under a block-encoding promise (Sec. 7).

The natural next steps follow from the construction itself: per-point cost accounting for the Lorentz- and real-part-kernel primitives and the adaptive window-selection protocol (Sec. 8.2 (iv)), an analytic bound on the negative moments, a sampling theory for METTS, a theorem-level treatment of the degree-sparing variant, and the higher-order construction, for which Sec. 4.4 supplies the route: the chain lives on the same $H_\Delta$ register, applied repeatedly.

## Data availability statement

The Python code, the exact-diagonalization module, archived result files, and figure-generation scripts are openly available at

https://gitee.com/xie-qingxing/thermal-window-weights

Results are keyed by the paper's check identifiers (e.g. V1, V4, T5k, tp6-C1), with a README mapping each figure and table to its result file. 

## Funding

This work is supported in part by the National Natural Science Foundation of China (Grant No. 12504267), the Natural Science Foundation of Hubei Province (Grant No. 2025AFB274).

## Competing interests

The authors declare no competing interests.